%% file: draft.tex
\documentclass[oldversion]{aa}  
\usepackage{graphicx}
\usepackage{txfonts}
\usepackage{natbib}
\bibpunct{(}{)}{;}{a}{}{,} 

\begin{document}
 
\title{
Strong Lensing in Abell~1703:\\ Constraints on the Slope of the Inner Dark Matter Distribution
}

   \subtitle{}

   \author{Marceau Limousin\inst{1,2}, Johan Richard\inst{3}, Jean-Paul Kneib\inst{4}, Henrik Brink\inst{2}, 
   Roser Pell\'o\inst{5},
   Eric Jullo\inst{4},
   Hong Tu\inst{6,7,8},\\
   Jesper Sommer-Larsen\inst{9,2},
   Eiichi Egami\inst{10},
   Micha{\l} J. Micha{\l}owski\inst{2},
   R\'emi Cabanac\inst{1}
   \& Daniel P. Stark\inst{3}
    }

   \offprints{marceau.limousin@ast.obs-mip.fr}

   \institute{
	Laboratoire d'Astrophysique de Toulouse-Tarbes, Universit\'e de Toulouse, CNRS,
	57 avenue d'Azereix, 65\,000 Tarbes, France
   	\and
         Dark Cosmology Centre, Niels Bohr Institute, University of Copenhagen,
        Juliane Marie Vej 30, 2100 Copenhagen, Denmark
	\and
	Department of Astronomy, California Institute of Technology, 105-24, Pasadena, CA91125, USA
	\and
	Laboratoire d'Astrophysique de Marseille, CNRS-Universit\'e de Provence, 38 rue Fr\'ed\'eric Joliot-Curie, F-13388 Marseille cedex 13, France
	\and
	Laboratoire d'Astrophysique de Toulouse-Tarbes, Universit\'e de Toulouse, CNRS,
	14 avenue Edouard Belin, 31\,400 Toulouse, France
	\and
	Physics Department, Shanghai Normal University, 100 Guilin Road, Shanghai 200234, China
	\and
	Institut d'Astrophysique de Paris, CNRS, 98bis Bvd Arago, 75014 Paris, France
	\and
	Shanghai Astronomical Observatory, 80 Nandan Road, Shanghai 200030, China
	\and
	Excellence Cluster Universe, Technische Universit\"at M\"unchen; Boltz-manstr. 2, D-85748 Garching, Germany
	\and
	Steward Observatory, University of Arizona, 933 North Cherry Avenue, Tucson, AZ 85721, USA
              }

   
  \abstract
   {
   Properties of dark matter haloes can be probed observationally and numerically, and comparing both
   approaches provide ways to constraint cosmological models.
   When it comes to the inner part of galaxy cluster scale haloes, interaction between the baryonic and
   the dark matter component is an important issue which is far to be fully understood.
   With this work, we aim to initiate a program coupling observational and numerical studies in order to
   probe the inner part of galaxy clusters. In this article, we apply strong lensing techniques in Abell~1703, 
   a massive X-ray luminous galaxy cluster at $z=0.28$. Our analysis is based on imaging data both from space 
   and ground in 8 bands, complemented with a spectroscopic survey.
   Abell~1703 is rather circular from the general shape of its multiply imaged systems and is dominated by a
   giant elliptical cD galaxy in its centre. This cluster exhibits a remarkable bright '\emph{central ring}'
   formed by 4 bright images at $z_{\rm spec}=0.888$ 
   only 5\,-\,13$\arcsec$ away from the cD centre. This unique feature offers a rare lensing constraint to probe the central mass distribution.
   The stellar contribution from the cD galaxy ($\sim$\,1.25\,10$^{12}$\,M$_{\sun}$ within 30\,kpc) is 
   accounted for in our parametric mass modelling, and the underlying
   smooth dark matter component distribution is described using a generalized \textsc{nfw} profile parametrized with a central
   logarithmic slope $\alpha$. 
   The \textsc{rms} of our mass model in the image plane is equal to 1.4$\arcsec$.
   We find that within the range where observational constraints are present
   (from $\sim\,20$\,kpc to $\sim\,210$\,kpc), $\alpha$ is
   equal to $1.09^{+0.05}_{-0.11}$ (3$\sigma$ confidence level).
   The concentration parameter is equal to $c_{200} \sim 3.5$, and the scale radius is constrained to be larger than
   the region where observational constraints are available ($r_s$\,=\,730$^{+15}_{-75}$\,kpc). 
   The 2D mass is equal to M\,(210\,kpc)\,=\,2.4\,10$^{14}$\,M$_{\sun}$.
   Although, we cannot draw any conclusions on cosmological models at this point since we lack results from realistic
   numerical simulations containing baryons to make a proper comparison.
   We advocate the need for a large sample of well observed (and well constrained)  and simulated unimodal relaxed galaxy clusters in order
   to make reliable comparisons, and potentially provide a test of cosmological models.
   }


    \titlerunning{Strong Lensing in Abell~1703}
   \authorrunning{Limousin et~al.}
   \maketitle
 
\section{Introduction}

Large N-body cosmological simulations have been carried out for a decade, with the goal of making
statistical predictions on dark matter (DM) halo properties.
Because of numerical issues, most of these large cosmological simulations contain dark matter particles only.
They all reliably predict that the 3D density profile $\rho_{\rm DM}(r)$ should fall
as $r^{-3}$ beyond what is usually called the scale radius. Observations have confirmed these predictions
\citep[\emph{e.g.}][]{kneib03,etienne,rachel08}.
This agreement is likely to be connected with the fact that at large radius, the density profile of a
galaxy cluster is dark matter dominated and the influence of baryons can be neglected.
On smaller scales, if we parametrize the 3D density profile of the DM using a cuspy profile $\rho_{\rm DM}\propto
r^{-\alpha}$; dark matter only simulations predict a logarithmic slope $\alpha \sim$ 1-1.5 for
$r\,\rightarrow$ 0. The exact value of the central slope and its universality is debated \citep{nfw,moore98,ghigna00,
ricotti,navarro04,gao07}.
From the theoretical  point of view, the logarithmic slope of DM halos is predicted to be $\alpha \sim$ 0.8 \citep{austin,steen}.
Although this debate is of some interest, these dark matter only studies and
their predictions
do not help much to make comparison with observations. Indeed, observing the central part
(i.e. the inner $\sim$ 500\,kpc) of a galaxy cluster at any wavelength reveals the presence of baryons (in the forms of stars and X-ray hot gas).
Thus any attempt to compare observations to simulations in the centre of galaxy clusters has to be
made with numerical simulations (or calculations) taking into account the baryonic component and its associated
physics.

Efforts are currently developed on the numerical side in order to account for the presence of baryons
(\emph{e.g.} the Horizon\footnote{http://www.projet-horizon.fr} simulation).
In practice, our understanding of the baryonic physics is
poor, and the exact interplay between dark matter and baryon is far from understood.
For example, due to the overcooling
problem, numerical simulations predict blue central brightest cluster galaxies which are not always observed.
Different effects do compete when it comes to the central slope of the density profile:
the cooling of gas in the centre of dark matter haloes is expected to lead to a more concentrated dark matter
distribution \citep[the so-called adiabatic contraction, see][]{blumenthal,gnedin,gustafsson}.
On the other hand, dynamical friction heating of massive galaxies against the diffuse cluster dark matter
can flatten the slope of the DM density profile \citep{elzant01,nipoti03,nipoti,MBK},
and this effect could even dominate over adiabatic contraction \citep{elzant04}.
Note also that the properties of the inner part of simulated galaxy clusters (even in dark matter only simulations)
can depend significantly on initial conditions as demonstrated in \cite{maxwellhalo}.
To summarize, no coherent picture has yet emerged from N-body simulations when it comes to the shape
of the inner density profile of structures; this problem is a difficult one and the answer is likely
not to be unique but may depend on the physical properties of the structures and their formation
history.

On the observational side, efforts have been put on probing the central slope $\alpha$ of the
underlying dark matter distribution. These analyses have led to wide-ranging results, whatever the method
used: X-ray
\citep[][]{ettori02,arabadjis02,lewis03,zappacosta06};
lensing \citep{tyson98,S01,dahle03,sand02,gavazzi03,gavazzi05,sand04,sand07,bradac} or dynamics 
\citep{kelson02,bivianosalucci}.
This highlight the difficulty of such studies and the possible large scatter in the value of $\alpha$
from one cluster to another.

To summarize, we need to probe observationally and numerically the behaviour of the \emph{underlying} dark matter 
distribution (i.e. after the baryonic component has been separated from the
dark matter component) in the central parts of galaxy clusters. 
The main difficulties are: i) Observationally, to be able to disentangle the baryonic component and the
underlying dark matter distribution;
ii) Numerically, to implement the baryonic physics into the simulations;
iii) Then to compare both approaches in a consistent way.
These issues are far beyond the scope of this article, but they are likely to provide an interesting test of the
$\Lambda$CDM scenario in the future.

\begin{figure*}[h!]
   \centering
\includegraphics[height=24.7cm,width=18cm]{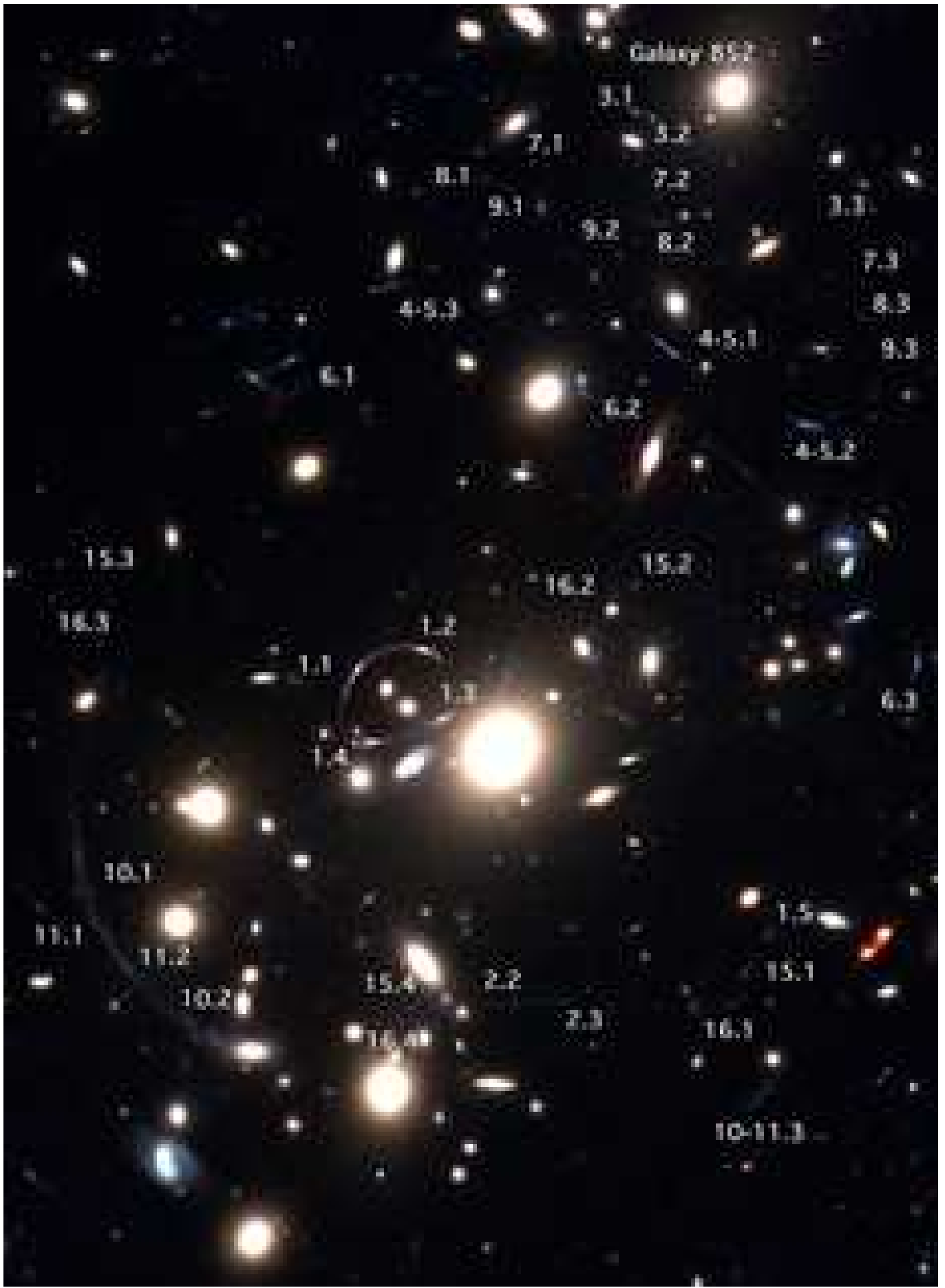}
    \caption{Colour image of Abell~1703 from F850W, F625W and F475W observations \citep{stott}.
    North is up, east is left. Size of the field of view is equal to
    $77\arcsec\, \times \, 107\arcsec$, corresponding to 326\,kpc $\times$ 454\,kpc.
    Multiply imaged systems used in the analysis are shown, and colour images are given in Appendix.
    The \emph{central ring} formed by four bright images is found close to the cD galaxy.
    The giant arc (systems 10-11) falls south-east, at distance of $\sim 35\arcsec$.
    System 2 is a straight arc located south of the cD and composed of two merging images.
    System 15 and 16 follow each other, forming a kind of Einstein cross configuration.
    In the north, we find a set of tangential systems (4-5-6-7-8-9).
    Then, two bright merging images form system 3, located close to galaxy 852 which
    present a blue nearby lensing feature.
    The filamentary structure can be appreciated on this image. See also the Subaru
    H band image in Appendix which is more extended.
    }
\label{nicefig}
\end{figure*}

In this work, we aim to probe observationally the central (i.e. from $\sim 5\arcsec$ ($\sim 20$\,kpc) up to
$\sim 50\arcsec$ ($\sim 200$\,kpc) from the centre) density profile of a massive cluster lens. 
Our main goal is to measure the slope of the inner underlying dark matter distribution within
this radius. In practice, we apply strong lensing techniques on galaxy cluster Abell~1703,
a massive $z=0.28$ \citep{allen92} X-ray cluster with a luminosity
L$_{\rm{x}}$ = 8.66\,10$^{44}$ erg\,s$^{-1}$ \citep{hans00}.
It is very well suited for the analysis we want to perform for the following reasons:
\begin{enumerate}
\item It contains a large number of gravitational arcs, providing many lensing constraints for the analysis (Fig.~\ref{nicefig}).
\item Although it displays an intriguing filamentary structure along the north-south direction,
it looks rather circular from the geometrical configuration of its multiply imaged systems,
in particular its giant arc, located at large angular separation ($\sim35\arcsec\sim147$\,kpc).
It is likely to be a unimodal cluster, with a clear dominant elliptical cD galaxy, which makes it much easier to interpret 
the results of the modelling compared to bimodal clusters such as Abell~1689 \citep{mypaperIII}, Abell~2218
\citep{ardis2218}, Abell~68 \citep{a68} or MS\,2053.7-0449 (Verdugo et~al., 2007).
Indeed, regular relaxed clusters are rare at such redshifts \citep{smith05}.
\item It presents a remarkable lensing configuration, forming a '\emph{central ring}' composed of four bright
images (see Fig.~\ref{nicefig} and Appendix).
Interestingly, these constraints are found very close to the cD galaxy, potentially providing robust 
constraints in the very central part of the cluster. 
Note that this '\emph{central ring}' is not really central since it does not coincide with the peak of
the total mass distribution.
\end{enumerate}

This article is organized as follows:
data used in this work are presented in Section 2.
Section 3 presents the multiply imaged systems that constitutes the basis of our analysis.
The strong lensing analysis is outlined in Section 4, and the results are given in Section 5.
We discuss our results and conclude in Section 6.

All our results are scaled to the flat,
$\Lambda$CDM cosmology with $\Omega_{\rm{M}} = 0.3, \
\Omega_\Lambda = 0.7$ and a Hubble constant \textsc{H}$_0 = 70$
km\,s$^{-1}$ Mpc$^{-1}$. In such a cosmology, at $z=0.28$, $1\arcsec$
corresponds to $4.244$ kpc. All the figures of the cluster are
aligned with WCS coordinates, i.e. north is up, east is left.
The reference centre of our analysis is fixed at the cD centre:
R.A.=13:15:05.276, Decl.=+51:49:02.85 (J\,2000.0).
Magnitudes are given in the AB system.
\clearpage
\include{table}

\section{Data}
Abell~1703 has been observed from space by the \emph{Hubble Space Telescope} (using ACS and NICMOS); 
from ground with the Subaru telescope
\citep[using the Multi-Object InfraRed Camera and Spectrograph, MOIRCS,][]{moircs} and with the Keck 
telescope \citep[using the Low Resolution Imaging Spectroscograph, LRIS,][]{lris}.
Table~\ref{obs} sums up the different dataset.
\begin{table}
\begin{center}
\begin{tabular}{ccc}
\hline
\smallskip
Instrument  &  Filter  &  Exp. time ($s$) \\
\hline
\smallskip
\smallskip
ACS    & F435W & 7\,050  \\
\smallskip
\smallskip
ACS    & F475W & 5\,564  \\
\smallskip
\smallskip
ACS    & F555W & 5\,564  \\
\smallskip
\smallskip
ACS    & F625W & 8\,494 \\
\smallskip
\smallskip
ACS    & F775W & 11\,128 \\
\smallskip
\smallskip
ACS    & F850W & 17\,800 \\
\smallskip
\smallskip
NICMOS  & F110W & 2\,624 \\
\smallskip
\smallskip
MOIRCS    & H & 18\,875 \\
\hline
\smallskip
\end{tabular}
\end{center}
\caption{Different observations of Abell~1703 used in this work: instrument, filter and exposure time in seconds.
}
\label{obs}
\end{table}

\subsection{Imaging Data}
The multiwavelength ACS data have been used to construct colour images of Abell~1703 in order 
to identify the mutiply imaged systems.
In addition to the 6 ACS bands, we benefited from NICMOS and Subaru data in order to construct
spectral energy distributions (SED) and estimate photometric redshifts for the multiple images as well as the 
stellar mass of the cD galaxy.
\paragraph{HST/ACS:}
We use the HST ACS data in six bands. Abell~1703 has been observed
on May 2005 as part of the ACS team guaranteed observing time (P.I. Holland Ford, proposal 10325).
These images have been reduced using the \emph{multidrizzle} software to
remove cosmic rays, bad pixels, combine the dithered frames and correct
for geometric distortions. The output pixel scale was fixed at
0.04\arcsec and we used a \emph{pixfrac} parameter value of 0.8 for reducing
the area of the input pixels.
\paragraph{HST/NICMOS:}
Abell~1703 has been observed with NICMOS on October 10, 2006 (P.I. Holland Ford,
proposal 10996), using the F110W filter.
NICMOS images have been reduced following the NICMOS data reduction
handbook\footnote{http://www.stsci.edu/hst/nicmos/documents/handbooks}
and including specific improvements to remove cosmic rays,
quadrant-to-quadrant variations, bias and flat residuals. More details on
these improvements are given in \citet{nicmos}.

\paragraph{Subaru MOIRCS:}
MOIRCS observations of Abell 1703  were obtained in May 2007 (PI: E. Egami).
We took five series of dithered exposures, with
individual exposure times of 20, 25 and 30 seconds and 5 coadds per frame,
producing a total exposure time of 18\,875 seconds in the central region.
These observations were obtained under excellent seeing conditions
(0.3-0.4\arcsec). We used the MCSRED
package\footnote{http://www.naoj.org/staff/ichi/MCSRED/mcsred.html} to perform
flat-fielding, sky subtraction, distortion correction and mosaicking of
individual images. The photometric calibration was derived with the
observed magnitudes of 15 stars from the 2MASS catalog.

\subsection{Spectroscopic Data}
We used LRIS on Keck I in an attempt
to measure a redshift for the brightest component of system 1 (image 1.1). Four exposures
of 900 seconds were taken on
Jan 29th 2008, under photometric conditions but a poor seeing
(1.4\arcsec), using the 900 lines.mm$^{-1}$ grating blazed at 6320 \AA\
in the red channel of the instrument. This setup covers the wavelength
range 5600-7200 \AA\ at a resolution of 2.77 \AA\ and a dispersion of 0.83
\AA\ per pixel.
The spectrum has been reduced with standard IRAF procedures for
bias correction, flat-fielding, sky subtraction and distortion correction,
in that order. We used the numerous sky lines in this region for the
wavelength calibration and observation of the standard star Feige 92 for
the flux calibration.

\begin{figure*}
\centering
\includegraphics[height=5.6cm,width=14.0cm]{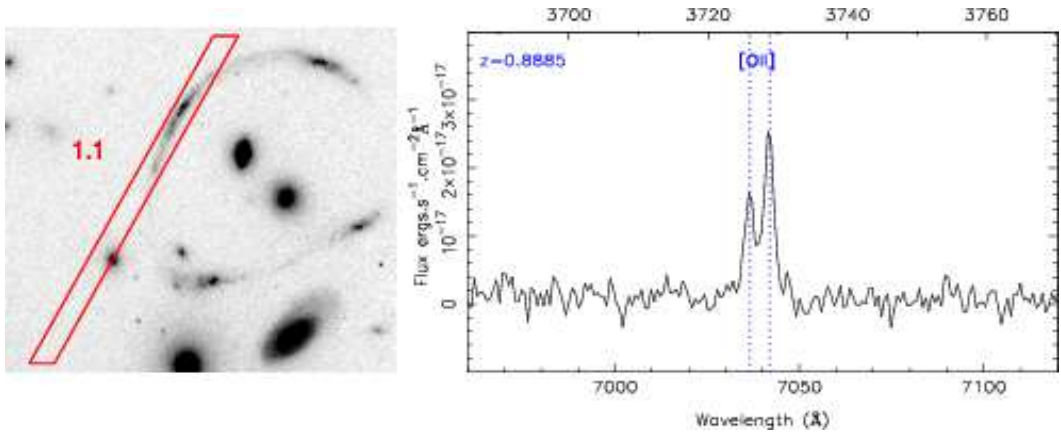}
\caption{
Spectroscopic observation of image 1.1.
\emph{Left:} Position of the slit. \emph{Right:} 1D spectrum. The $[O_{II}]_{3726,3729}$
doublet is easily identified.
}
\label{zspec}
\end{figure*}

We detected a bright doublet of emission lines, centered at 7036 and 7041
\AA\ respectively, which we interpret as $[O_{II}]_{3726,3729}$ at
$z=0.8885\pm0.0002$ without any ambiguity, the doublet being easily
separated by 5$\AA$ at this resolution (Fig.~\ref{zspec}).
The spectroscopic
redshift is in agreement with our photometric redshift estimate
of $z_{\rm phot}=0.965^{+0.075}_{-0.240}$ (Table~\ref{multipletable}).

\subsection{Photometric Analysis}
We created photometric catalogs combining the multicolour images by running
SExtractor \citep{sextractor} in \textit{double image} mode. A specific detection image
was created by combining all the ACS data after scaling them to the same
background noise level. We used the cD-subtracted images (see Section~\ref{cDgalax}) for ACS and
MOIRCS to prevent any strong photometric contamination in the central
regions, in particular for the measurement of the lensed ring-shape
system (system 1).
Photometry was optimized to the small size of the lensed background
galaxies, by measuring the flux in a
1.0\arcsec diameter aperture. This gives accurate colours in the optical
bands that used the same instrument. We estimated aperture corrections
in the near-infrared by measuring the photometry of 10 isolated bright
point sources in the NICMOS and MOIRCS images, and corrected our
photometric catalogs for this difference.

We increased the photometric error bars, usually underestimated by
SExtractor, to take
into account the effects of drizzling in the reduction of ACS and NICMOS
images, following the computations by \citet{casertano}. For the
MOIRCS images, we measured the pixel-to-pixel background noise from blank
regions of sky selected in the original images, and scaled it to the
aperture size used in the photometry.

\subsection{Cluster Member Identification}
To extract cluster galaxies, we plot the characteristic cluster red sequences (F775W\,-\,H) and (F625W\,-\,H) in two
colour-magnitude diagrams and select the objects lying on both red-sequences as cluster galaxies.
This yields 345 early-type cluster galaxies down to F775W\,=\,24.
For the purpose of the modelling, we will consider only the cluster members whose
magnitude is brighter than 21 in the F775W band \citep[in order to save computing time, see][]{ardis2218} and 
which are located close to some multiply imaged systems.
We consider fainter galaxies only if they are located close to some
multiple images since they can locally perturb the lensing configuration \citep{arcsubstructure}.
This yields 45 galaxy scale perturbers.

\subsection{Photometric redshifts}
We ran the photometric redshift code \textit{HyperZ} \citep{hyperz} on the multi-band 
photometric catalogs, in order to get a redshift
estimate for all the multiple images identified in the field. This
program performs a minimization procedure between the spectral
energy distribution of each object and a library of spectral
templates, either empirical \citep{Kinney,CWW} or
from the evolutionary models by \citet{BC03}. 
Note that the NICMOS data we have correspond to a mosaic of four pointings, none of them being centred on the
cD galaxy. It results that the cD galaxy is not fully sampled by the NICMOS data.
Therefore, it cannot be subtracted in the NICMOS image, thus we did not take into
account the F110W filter measurement for all its neighbouring objects, or
for multiple images located at the edges of the NICMOS field of view.
Absolute photometric calibration between the different bands is usually
accurate down to 0.05 magnitudes, we used this value as a minimal value
in the code (this value will be used also for fitting the SED of the cD galaxy in Section 2.6). 
We scanned the following range of
parameters: $0.0<z<7.0$ for the redshift, $0.0<A_V<1.2$ for the
reddening, applied on the template spectra with the \citet{calzetti}
law observed in starburst galaxies. Optical depth in
the Lyman-$\alpha$ forest followed the \citet{madau95} prescription.

The results obtained for each multiple image are reported in Table~\ref{multipletable},
along with the 3$\sigma$ error bar estimate given by \textit{HyperZ} from the
redshift probability distribution.

\subsection{cD galaxy}
\label{cDgalax}
Abell~1703 exhibits a dominant central giant elliptical galaxy: its stellar contribution to the mass budget
in the central part is to be taken properly into account.
We worked out some of the cD galaxy properties from the broad band photometry.
In particular, we are interested in computing its luminosities in the different filters in order to estimate its
stellar mass.
Note again that we have not been able to study the cD galaxy in the NICMOS band since 
the NICMOS data correspond to a mosaic of four pointings, none of them being centred on the
cD galaxy.

\begin{figure}
\centering
\includegraphics[height=7cm,width=7cm]{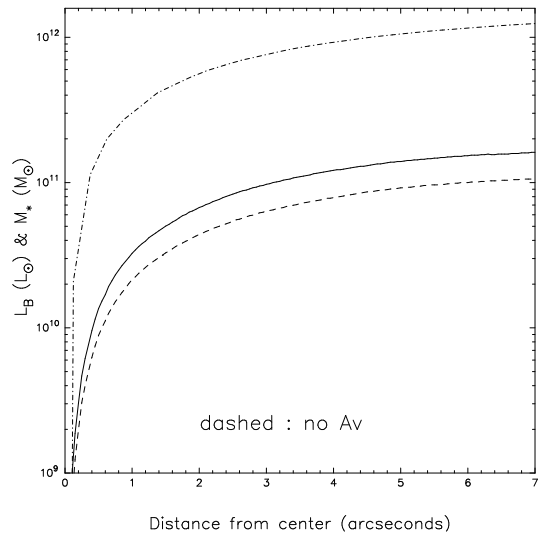}
\caption{
Integrated luminosity profile of the cD galaxy, in the B band rest frame, for the fit with redenning (solid) and
without redenning (dashed),
The corresponding mass profile used in the strong lensing analysis is shown as dot-dashed line.
Both profiles have the same behaviour for
R\,$>\,\sim2\arcsec$.
}
\label{M_L}
\end{figure}

\paragraph{Subtraction of the cD galaxy:}
We fitted and subtracted from each image a model representation
of the surface brightness distribution using the IRAF task \emph{ellipse}. Both the position angle 
and ellipticity were allowed to vary as a function of the semimajor axis in the fitted elliptical
isophotes, as well as the isophote centroid in the central part. This procedure was found to give
satisfactory residuals at the centre (see Appendix).
This procedure also allows us to determine accurate photometry for the cD galaxy, that we 
report in Table~\ref{cD}.
The integrated luminosity profile in the rest frame B band
is shown in Fig.~\ref{M_L}, together with a mass profile of the cD galaxy
as included in the modelling (see Section~\ref{cDmodel}).

\begin{figure}
\centering
\includegraphics[height=8cm,width=6cm,angle=-90]{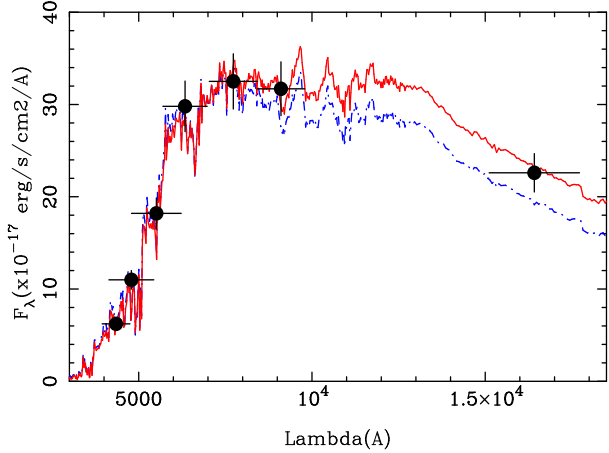}
\caption{
Results of the SED fitting procedure. Luminosities are computed in an aperture of 7$\arcsec$.
The red solid line corresponds to the fit wit redenning, and the blue dotted-dashed line corresponds to the fit 
without redenning.
In each case, the fit is good ($\chi^2_{\rm d.o.f.}<\,1$) and leads to the same estimation of the
stellar mass.
}
\label{SED}
\end{figure}
\begin{table}
\begin{center}
\begin{tabular}{ccccc}
\hline
\smallskip
Filter  &  L (no Av) & L (Av) & M$_{*}/$L (no Av) & M$_{*}/$L (Av)\\
\hline
\smallskip
\smallskip
B    & 1.1\,10$^{11}$ &1.6\,10$^{11}$ & 11.2 & 8.1 \\
\smallskip
\smallskip
555  & 1.5\,10$^{11}$ & 2.2\,10$^{11}$ & 7.8 & 5.9 \\
\smallskip
\smallskip
606  & 1.7\,10$^{11}$ & 2.5\,10$^{11}$ & 6.8 & 5.2 \\
\smallskip
\smallskip
775  & 1.5\,10$^{11}$ & 2.2\,10$^{11}$ & 7.7 & 6.0 \\
\smallskip
\smallskip
H  & 7.2\,10$^{11}$ & 9.5\,10$^{11}$ & 1.6 &  1.4\\
\hline
\end{tabular}
\end{center}
\caption{
Luminosities and stellar mass to light ratio computed in different rest frame filters.
The B filter corresponds to the standard Bessell B filter, and values are given in this filter
in order to allow easier comparisons with other galaxies studied by different authors.
All values are given in solar units.
We report values corresponding to two different fits, with and without reddening respectively.
}
\label{cD}
\end{table}

\paragraph{Stellar mass:}
From the broad band photometry, we estimate the stellar mass of the cD galaxy by fitting
its spectral energy distribution using the \textit{HyperZ} software.
In order to perform this estimation, we choose an aperture equal to 7$\arcsec$.
This choice is motivated by the goodness of the fit in this aperture ($\chi^2_{\rm d.o.f.}<\,1$).
Moreover, we find the
stellar mass to light ratios (M$_{*}/$L) to be constant within this aperture.
We use 8 evolutionary synthetic SEDs computed with the last version of the \citet{BC03} code, with
\citet{chabrier} IMF and solar metallicity.
Given the data available, the photometric SED is equally well fit with either a very old elliptical template
(a Single Stellar Population, SSP, aged 10.2 Gyr), or a slightly younger population
(SSP aged 6-7 Gyr) including some reddening (A$_{\rm V}$=0.4, with \citet{calzetti} extinction law).
The degeneracy in the (A$_{\rm V}$, age) plane yields slightly different results in the best fit template
depending on the aperture radius. However, both types of models (i.e. with and without reddening) are
equally likely, as shown in Fig.~\ref{SED} for the adopted 7$\arcsec$ aperture. Although the total 
luminosity and M$_{*}/$L ratio depend on the reddening correction, the stellar mass within a given
aperture is found to be insensitive to the best fit template.
The resulting stellar mass within a 7$\arcsec$ aperture is M$_* = 1.25\pm0.3\,10^{12}\,$M$_{\sun}$.
According to \citet{bell}, the errors coming from setting the overall mass scale and its evolution are
usually 25\%, much larger than the error on the photometry. We adopt 25\% accuracy on the stellar mass
estimation.
Given the luminosity of this galaxy, we derive a
a stellar mass to light ratio in the rest frame B band equal to $\sim 8-11$ (solar units), depending on the
redenning. We report all these values in Table~\ref{cD}.
All luminosities are given in solar units calculated in the rest frame filters.

This is comparable (though a bit higher, but note that this galaxy has a very massive stellar population) 
to typical values of stellar mass to light ratio for giant elliptical 
galaxies \citep{gerard01}.

\section{Multiply Imaged Systems}
Fig.~\ref{nicefig} shows a colour image of Abell~1703 where we label the multiply imaged systems.
We report their positions and photometric redshifts in Table~\ref{multipletable}.
The identification is a difficult step which is done in an iterative fashion: we begin
to build a model using the most obvious lensed features. Then this model is used
to test and predict possible multiply imaged systems.
In total, we use 13 multiply imaged systems in this analysis.
It is certain that more systems have to be found within the ACS field that presents
many likely blue lensed features.

Here we give some notes on the different systems, and present colour images for each system in Appendix.
\begin{itemize}
\item System 1, the '\emph{central ring}':
it is composed by 5 images. The four main bright images located close to the cD
galaxy display a rare ``hyperbolic umbilic'' lensing configuration (see Fig.~\ref{nicefig} and Appendix).
A demagnified counter image is located in the east.
Image 1.3 is located at 5.5$\arcsec$\,=\,23\,kpc from the centre of the cluster,
whereas image 1.1 is located at 13.5$\arcsec$\,=\,56\,kpc from the cluster centre.
This system constitutes the innermost observational lensing constraint available in this analysis.
This is the only system for which we have a spectroscopic redshift, equal to
0.8885.
\item System 2:
This system is constituted by two images forming a straight gravitational
arc. Two counter images are predicted to be more than two magnitudes fainter, we have not been able
to detect any.
\item System 3:
System 3 is located in the northern part of the ACS field, close to Galaxy 852.
It is composed of two merging bright images, and an additional fainter counter image
a bit further west.

\item System 4-5:
We propose that a single lensed galaxy can be resolved into two parts, each part
being considered as a multiply imaged system.
It constitutes a typical cusp configuration system, with the images forming only on one side of the cluster.
\item System 6:
This is another cusp configuration system located a bit closer to the centre than system
4-5.
\item Systems 7-8-9:
These three cusp configuration systems are located in the north, at a radius a bit further from the centre 
than system 4-5.
\item System 10-11:
These two systems constitute the giant tangential arc.
They correspond to two different spots we have identified on the giant arc
that shows many substructures.
\item System 15 and 16:
These two systems composed by 4 images 'follow' each other (Fig.~\ref{nicefig}):
15.$i$ and 16.$i$ ($i$=1,2,3,4) are found close to each other, presenting an Einstein cross
configuration.
\item Radial Arc:
We report a radial feature coming out from the cD galaxy, composed by two spots (Fig.~\ref{sys1}).
Due to the presence of the cD, the estimated photometric redshift ($z\sim 1.4$) is uncertain, and we have not
been able to detect possible counterimages. This likely radial arc is not used in the analysis however. 
\end{itemize}

\newpage
\section{Strong Lensing Analysis}

\subsection{Methodology}
To reconstruct the mass distribution in Abell~1703, we use a parametric method
as implemented in the publicly available \textsc{lenstool}\footnote{http://www.oamp.fr/cosmology/lenstool/}
software \citep{jullo07}.
We use the observational constraints (positions of the multiply imaged systems) to
optimize the parameters used to describe the mass distribution: this
is what we refer to as optimization procedure.
The strong lensing methodology used in this analysis has been described in details in \citet{mypaperIII}.
We refer the interested reader to this article for a complete description of our methodology.

We describe the mass distribution in Abell~1703 by constructing a two components mass model:
the contribution from the dominant central cD galaxy is fixed by its stellar mass, and then the remaining
mass is put into an underlying smooth dark matter distribution described using a generalized
NFW profile \citep[see][for details on the implementation of the generalized NFW profile into the
\textsc{lenstool} code]{sand07}. We also take into account the perturbations associated with the galaxies.

\subsection{Modelling the cD contribution}
\label{cDmodel}
Degeneracies can arise between the two mass components: if too much mass is put into the cD galaxy, then
this can lead to a shallower slope of the dark matter halo and vice versa.
Therefore, special care has to be taken when modelling the cD galaxy, and such a modelling must be as much as possible
'observationally motivated'.
We use a dual Pseudo Isothermal Elliptical Mass Distribution \citep[dPIE, see][]{ardis2218} with no
core radius to describe the cD stellar mass contribution.
This profile is formally the same as the Pseudo Isothermal Elliptical Mass Distribution (PIEMD) profile
described in \citet{mypaperI}.
However, as explained in \citet{ardis2218} it is not the same as the PIEMD originally defined by
\citet{kassiola}. Therefore we have adopted the new name dPIE to avoid confusion.
The position of this mass clump is fixed at (0,\,0); the ellipticity and position angle
are set to be the one we measured from the light distribution.
Given this parametrization, the mass of the cD scales as M$_{\rm{cD}}\,\propto \sigma_0^2 \times r_{\rm{cut}}$, where 
the scale radius $r_{\rm{cut}}$ is almost equal to the half mass radius, and $\sigma_0^2$ is a fiducial
velocity dispersion \citep{ardis2218}.
The choice of the scale radius will influence the shape of the mass profile.
The smaller the scale radius, the steeper the mass profile.
We choose a scale radius of 25\,kpc so that the mass profile is as close as possible to the
integrated luminosity profile. As we can see on Fig.~\ref{M_L}, the mass profile and the luminosity
profile have the same behaviour for R\,$>\,\sim2\arcsec$.
Note that 25\,kpc is close to the value used by \citet{sand07} to describe the cD galaxies in 
MS\,2137 (22\,kpc) and Abell~383 (26\,kpc), where the same mass profile has been used.
The only free parameter describing this cD galaxy is then $\sigma_0$.
This parameter is allowed to vary in a range that is set by the choice of the scale radius and the 
estimation of the stellar mass.

\subsection{Modelling the underlying dark matter distribution}
We \emph{assume} that the dark matter component can be described using a generalized NFW model.
Its 3D mass density profile is given by:
\begin{equation}
\rho(r)=\frac{\rho_{c} \delta_{c}}{(r/r_{s})^{\alpha}(1+(r/r_{s}))^{3-\alpha}},
\end{equation}
with $r_s$ the scale radius (note that the scale radius for an NFW profile does not have the same meaning as the scale
radius for a dPIE profile, see \citet{mypaperI}).
For a galaxy cluster, this parameter is supposed to be larger than 150\,kpc (Section 4.5),
so in the range in radius we probe in this work, we have r$<r_s$ and thus the 
density profile can be approximated by $\rho(r) \sim r^{-\alpha}$.

Note that we have subtracted \emph{only} the stellar contribution of the cD galaxy in our modelling.
This is consistent with the general picture that the dark matter halo of the cluster is also the one of the
cD galaxy \citep[see, \emph{e.g.}][]{miralda95}.
We cannot distinguish both haloes.

\subsection{Galaxy scale perturbers}
On top of these two components, we include the brightest cluster members in the optimization (Section 2.3).
This galaxy scale component is incorporated into the modelling using empirical scaling relations that relate
their dynamical parameters (central velocity dispersion and scale radius) to their luminosity, whereas their
geometrical parameters (centre, ellipticity, position angle) are set to the one measured from the light
distribution \citep[see][for details]{mypaperIII}.
This galaxy scale component is thus parametrized by only two free parameters, and at the end of the optimization
procedure, we get constraints on the parameters for a galaxy of a given (arbitrary) luminosity which corresponds to an observed magnitude $m_{F775W}=18.3$.

One single galaxy (labeled 852 in our catalogue) is optimized individually, in the sense that some of the parameters
describing this galaxy are allowed to vary instead of being fixed by its luminosity.
This was necessary to reproduce better the geometrical configuration of some images falling close to this galaxy
(systems 3, 7, 8 and 9). In fact, a visual inspection at the image of the cluster shows that
this galaxy is very bright and extended (Fig.~\ref{nicefig}). This galaxy is not representative of the cluster 
galaxy population, thus the adopted scaling laws might not apply to this object.
Moreover, as shown in Appendix, a likely lensed blue feature is coming out from this galaxy,
suggesting a massive substructure.

\subsection{Limits on the parameters}
Each parameter is allowed to vary between some limits (priors).
The position of the DM clump was allowed to vary between $\pm\, 25\arcsec$ along the X and Y directions.
Its ellipticity was forced to be smaller than 0.5, since beyond that, the ellipticity as defined in this work
for an NFW profile is no longer valid \citep{golsenfw}.
Its slope $\alpha$ was allowed to vary between 0.2 and 2.0 and its scale radius $r_s$ between
150 and 750 kpc \citep{limitonrs,dolag04}.
The concentration parameter $c_{200}$ was allowed to vary between 2.5 and 9, which is large enought to include the
expectations from the most recent results from N-body simulations \citep{neto}.
The mass of the cD component was forced to be within the range allowed by the stellar mass estimate.
Concerning the galaxy scale component, we allowed the velocity dispersion
to vary between 150 and 250 km\,s$^{-1}$,
and the scale radius was forced to be smaller than 70 kpc, since we have evidences both from observations
\citep{Priya1,geigeramas,Priya2,Priya3,mypaperII,aleksi} and from numerical simulations \citep{mypaperIV} that
dark matter haloes of cluster galaxies are compact due to tidal stripping.

\section{Results}
\subsection{Mass Distribution from Strong Lensing}
Results of the optimization\footnote{A
parameter file containing all the following information, and which can
be used with the publicly available \textsc{lenstool} software, is
available at  http://www.dark-cosmology.dk/archive/.
This file can be useful for
making model based predictions, e.g. counter-images of a multiple
image candidate, amplification and mass map and location of the
critical lines at a given redshift, and it will be updated.}
are given in Table~4.
The images are well reproduced by our mass model, with an image  plane RMS equal to 
1.4$\arcsec$ (0.2$\arcsec$ in the source plane).
RMS for individual systems are listed in Table~\ref{multipletable}.
We derive from our model a 2D projected mass within 50$\arcsec$ equal to
M($50\arcsec$)\,=\,2.4\,10$^{14}$\,M$_{\sun}$.

\paragraph{Dark Matter Component:}
The logarithmic slope of the 3D dark matter distribution is found to be equal to  
$1.09^{+0.05}_{-0.11}$ (3$\sigma$ confidence level). 
We show in Fig.~\ref{res} degeneracy plots between $\alpha$
and: the mass of the cD galaxy, the concentration parameter $c_{200}$, and the scale radius $r_s$.
We find $c_{200}\sim [3.0-4.2]$ and M$_{200}\sim 1.8\,10^{15}$\,M$_{\sun}$. However,
we caution that these values rely on pure extrapolation from the strong lensing fit and that these
quantities should be probed using weak lensing and/or X-ray data.
Indeed, the NFW scale radius is found to be larger than the radius over which we have observational 
constraints (i.e. $\sim 54\arcsec$).
The ellipticity of the mass distribution is small, with $a/b=1.13$.
The location of the DM clump coincide with the position of the cD galaxy.
No second large scale dark matter clump was needed by the data. From previous experience modelling bimodal
galaxy clusters such as Abell~1689, Abell~2218, Abell~68 and MS\,2053.7-0449, we are pretty confident that there is
no need for a second large scale DM clump.

\paragraph{Galaxy Scale perturbers:}
The parameters inferred for the galaxy population (for an observed magnitude equal to 18.3 in the F775 band) are:
$\sigma_0=210\pm4$\,km\,s$^{-1}$, and $r_{\mathrm{cut}}\,=66\pm5$\,kpc.
We checked that the degeneracies between these two parameters
are the expected ones, in the sense that they follow constant mass lines 
(M\,$\propto \sigma_0 \times r_{\mathrm{cut}})$.
Galaxy 852 is found to be quite massive, with M\,(15$\arcsec$)$\sim$ 5\,10$^{12}$\,M$_{\sun}$.

\subsection{Mass \& Light: A Relaxed Unimodal Cluster ?}

We compare the mass and the light distribution in Fig.~\ref{massLight}. We find they do compare
well, suggesting that light traces mass in Abell~1703.
Also shown is the light distribution from the dominant cD galaxy, which is found to be consistent with the
one of the overall mass distribution within a few degrees.
Moreover, we find the centre of the DM halo to be coincident with the centre of the cD galaxy. 

At first approximation, these facts suggest that Abell~1703 is a relaxed unimodal cluster.
This hypothesis should be investigated further, in particular with precise X-ray observations
since we expect the X-ray emission to be centred on the cD galaxy, and to present a small ellipticity.

Besides, we can draw a line from the southern part of the ACS field
up to its northern part that pass through the cD galaxy and that connects very bright galaxies, much brighter than
the overall galaxy population (Fig.~\ref{nicefig}).
Interestingly, when looking at Abell~1703 on much larger scales from SDSS imaging 
(which size is $581\arcsec\,\times\,809\arcsec$),
the whole cluster galaxy population looks rather homogeneous, with no very luminous galaxies as we can observe
along this filamentary structure.
In the south of the ACS field, this structure seem to have an influence on the formation
of the giant arc, breaking its symmetry (Appendix).
The perturbation associated with the northern part of this structure is more obvious to detect.
Indeed, the formation of systems 3, 7, 8, and 9 (both their existence and their geometrical configuration)
is connected with this extra mass component. Moreover, we can see that a blue lensed feature
is coming out from Galaxy 852 (Appendix). The fact that we had to constrain individually Galaxy 852
also points out that some extra mass is needed in this region and that this galaxy (and possibly the other
bright galaxies defined by this filament) is not representative of the overall cluster population.
One tentative explanation could be that we are observing a galaxy group infalling in the cluster centre.
Though this scenario would need a devoted spectroscopic follow up of the cluster members to get
some insights into the velocity dimension of Abell~1703.

\subsection{Reliability of the constraints on $\alpha$}
One of the main goals of this work is to measure the slope of the underlying dark matter component,
parametrized by $\alpha$.
We want to stress out again that we have assumed that the underlying mass distribution can be
described using a generalized NFW profile, but this assumption may not be correct.
We try some tests in order to check the reliability of our measurement.
\paragraph{cD modelling:}
As discussed before, $\alpha$ will be sensitive to the parametrization of the cD galaxy.
During the first steps of this work, we noticed that if too much mass is put into the 
cD component, then $\alpha$ can become smaller than what is found here, and vice versa: if not enough
mass is put into the cD component, $\alpha$ can become larger.
However, these scenarios were leading to stellar masses that are not compatible with the photometry. 
Remain the choice of the scale radius of the cD galaxy. The value used in this analysis is set by the
shape of the integrated luminosity profile. We tried to use smaller (down to 10\,kpc) and larger 
(up to 45\,kpc) values of the scale radius, tuning up or down the velocity dispersion to keep the
stellar mass of the cD consistent with what we have estimated from the broad band photometry.
We find this does not have strong influence on the estimation of $\alpha$.
More precisely, when using a scale radius of 10\,kpc, we find $\alpha=1.10\pm0.04$, whereas when using
a scale radius of 45\,kpc, we find $\alpha=1.05\pm0.05$.

\paragraph{System 1:}
system 1 constitutes the innermost observational constraint in our analysis.
The configuration of the brightest 4 images forming the '\emph{central ring}' allows us
to probe the potential at small radii and this system constitutes our
most stringent constrain: if removing system 1 from the analysis, we found that the slope
was basically unconstrained, since we could get an equal fit with $\alpha\,\sim 0.5$ or $\alpha\,\sim 1.5$,
suggesting that the use of a generalized NFW profile was not necessary.
We note that the RMS of system 1 is slightly larger than the mean total RMS. This interesting lensing
feature is not perfectly retrieved in our analysis.
One could try to improve the situation by considering the influence of galactic substructures on the 
formation of this ring. However, from the large separation of the counter image 1.5, it is clear
that this central ring is not a galaxy-scale lens.
System 1 is also the only system for which we have been able to measure a spectroscopic
redshift. This allows us to check how the results presented in this work are dependent of the redshift of this
system. Instead of fixing its redshift to the measured one, $z=0.8885$, we
allowed it to vary between 0.72 and 1.0 as constrained from the photometry.
In this case, the model slightly underestimate the redshift and find $z\sim0.78$.
The slope of the underlying dark matter halo shifts to higher values, $\alpha\sim1.3$.
This highlight the importance of having spectroscopic redshifts over photometric redshifts when it comes
to detailed study of mass distributions.

\paragraph{Photometric redshifts:}
As we can appreciate in Table~1, most of the photometric redshifts are well constrained by the available
filters. In particular, the redshift probability functions were dominated by a well defined peak. 
In order to investigate the importance of using reliable photometric redshifts, we
redid the analysis as follows: the redshift of system 1 was fixed to the
spectroscopically measured value, and all other redshifts where assigned
a flat prior between 0.28 (redshift of the cluster) and 6.
Results of the optimized redshifts ($z_{\rm free}$) are given in Table~1.
We can see that they differ significantly from their photometrically constrained values, 
and that they are systematically overestimated.
Results on the parameters of the dark matter clump are the following:
$\alpha=1.22^{+0.03}_{-0.05}$; $c_{200}=3.4\pm0.3$; $r_s \sim 520$\,kpc.
Therefore, mainly the value of the slope is significantly changed when letting redshifts
being free. Since increasing $\alpha$ will lead to decreasing the mass
enclosed in a given radius, redshifts are being overestimated in order to compensate for. This highlight the importance of photometric redshift when spectroscopic redshift are not available.

\paragraph{X-ray gas component:}
In this analysis, we have not been able to subtract the mass contribution from the X-ray gas.
Recent analysis by \citet{bradac} in RXJ\,1347.5-1145, combining lensing and X-ray, has been able
to disentangle each component.
Indeed, they found that the total mass profile (DM+gas+stars) is slightly higher than the DM mass profile, 
the ratio of both quantities at 75\,kpc being equal to 85\%.
This could suggest that one can neglect the gas component at first approximation.
Of course, Abell~1703 and RXJ\,1347.5-1145 look very different thus we cannot compare them quantitatively.
To be more precise, we would need devoted X-ray observations of Abell~1703.

\paragraph{Limits on the parameters:}
The limits adopted in this work are motivated by results from 
numerical simulations. Here we want to investigate further what is the
influence of the adopted limits on the results, in particular on the slope
of the dark matter distribution.
First, we released these limits for $c_{200}$. 
When allowed to vary between 1 and 13 (instead of 2.5 and 9), we find the
parameters of the dark matter distribution to be fully consistent
with the results given in Table~4.

Another concern is the scale radius $r_s$. We find it to be larger than the
range within which observational constraints are available.
Since degeneracies arise between the scale radius and the slope of the
generalized NFW profile, we redid the analysis by imposing the scale radius
to be within different limits in order to see the influence on the results. 
Since we do have observational constraints up to 210\,kpc from the centre of the cluster, we are confident
that if $r_s$ was below 200\,kpc, our analysis would have been able to constrain it. Therefore, we
consider the following ranges: (1) 200-300\,kpc; (2) 300-400\,kpc; (3) 400-600\,kpc
and (4) 600-800\,kpc.
We report the results in Table~5:
(0) corresponds to the mass model presented in Section~5.1 and Table~4.
The following lines correspond to models obtained when assuming a different range of limits for the scale radius.
For each model ($i$), we report the inferred values for ($r_s,\,c_{200}\,,\alpha$) and compare each run with model (0).
In order to quantify this comparison, we report $\Delta$(log(Ev)) which is the difference between the Bayesian
Evidence of model (0) and the Bayesian Evidence of the considered model. When this quantity is positive, it
favours model (0) as being more likely.
We also report $\Delta(\chi^2)$ which is the difference between the $\chi^2$ of model (0) and the $\chi^2$ of the
considered model. When this quantity is negative, it favours model (0) as being a better fit to the data.

We can draw the following conclusions:
for models (1), (2) and (3), we see that the scale radius is always found at the higher end of the allowed
limit, suggesting that this parameter could be larger than the upper limit assigned.
As a result, we find these models to be less likely than the model (0), since both their $\chi^2$ and
Bayesian Evidence are worse.
The results for model (4) are essentially the same as for model (0): the parameters, as well as the $\chi^2$ are found
to be fully consistent. The Bayesian Evidence favours model (4) over model (0), which can be understood by the fact
that the ratio between the posterior and the prior is smaller in the case of model (4). 
Note that model (3) is also consistent with model (0), in the sense that the mass clump parameters
agree with each other, and that both Evidences and $\chi^2$ are comparable.

\begin{figure}
\centering
\includegraphics[height=8cm,width=8cm]{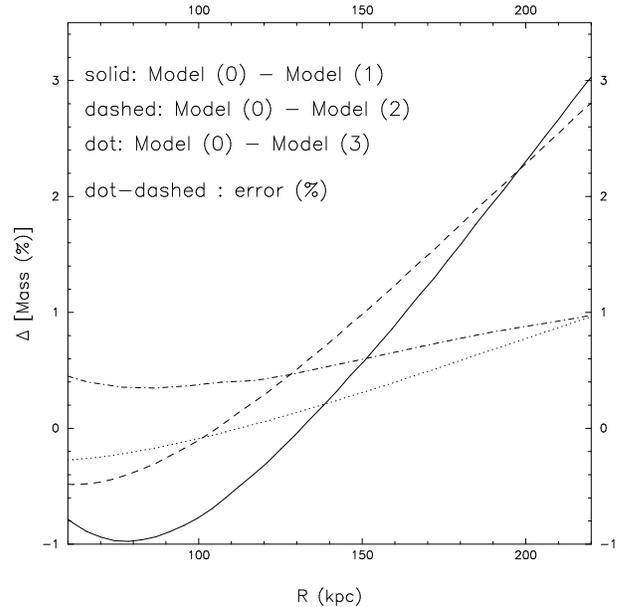}
\caption{
Mass differences between model (0) and model (1) (solid), model (2) (dashed) and model (3) (dotted).
The dot-dashed line correspond to the accuracy on the mass measurement, expressed as a percentage.
If mass differences are below the accuracy up to $\sim$\,125\,kpc, we see that
for R$>$150\,kpc, mass differences between model (0) and models (1) and (2) are above the accuracy on 
the mass measurement, which means that
we are able to discriminate between these different mass models.
}
\label{diffmass}
\end{figure}

\begin{figure}
\centering
\includegraphics[height=8cm,width=8cm]{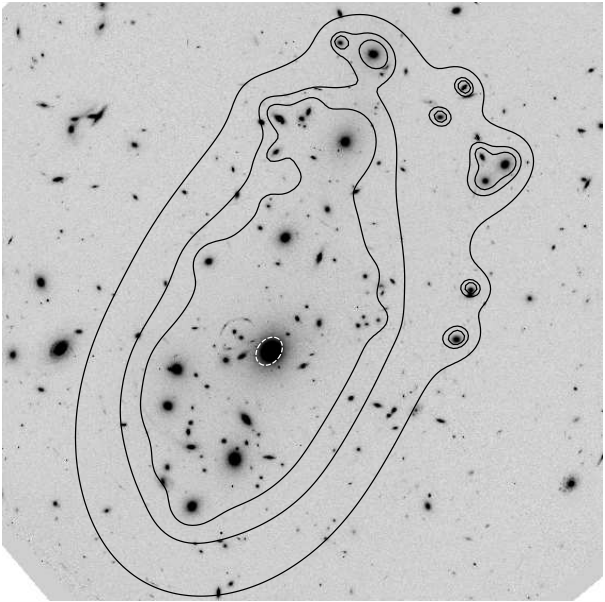}
\caption{ 
Mass contours overlaid on the ACS F850W frame. 
Also shown is the light distribution from the cD galaxy (dashed white contours), which is found to be consistent in orientation
with the one of the mass distribution within a few degrees.
Note how the light and the mass follow the filamentary structure.
Size of panel is $153\arcsec\,\times\,153\arcsec$.
}
\label{massLight}
\end{figure}

\begin{figure*}
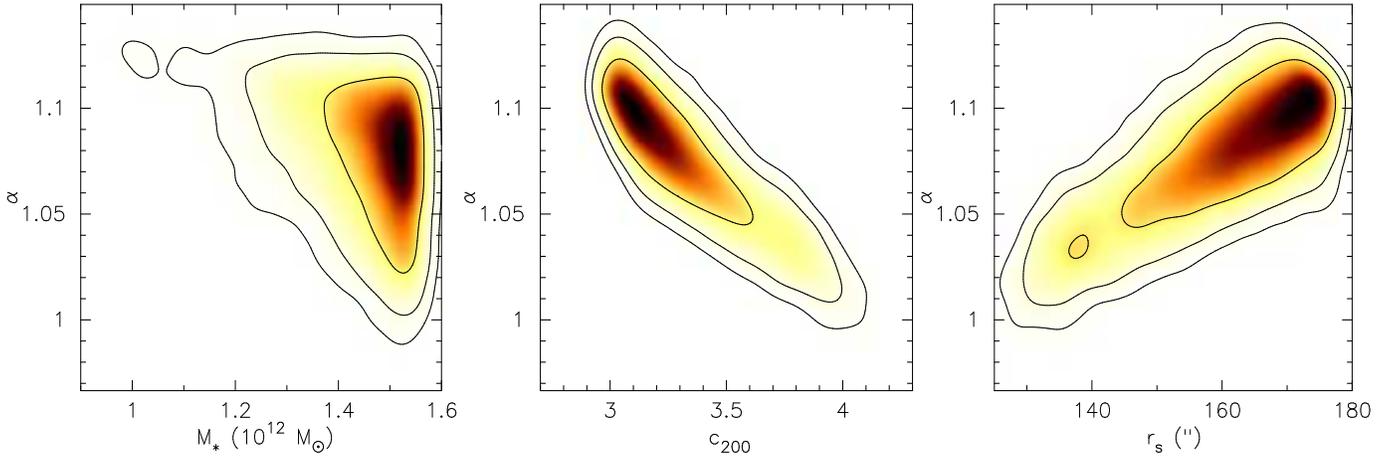

\centering
\includegraphics[height=6cm,width=6cm]{f6.1.ps}
\includegraphics[height=6cm,width=6cm]{f6.2.ps}
\includegraphics[height=6cm,width=6cm]{f6.3.ps}
\caption{
Degeneracy plots between $\alpha$ and (\emph{from left to
right}): cD stellar mass, $c_{200}$ and $r_s$.
$\alpha$ corresponds to the logarithmic slope of the underlying dark matter mass distribution
parametrized by a generalized NFW profile. $c_{200}$ and $r_s$ correspond to the concentration parameter
and the scale radius respectively for this mass distribution.
Note that the scale radius is found larger than the range within which we have observational 
constraints.}
\label{res}
\end{figure*}

These results suggest that, even though the preferred value for $r_s$ is found larger than the range over which
observational constraints are found, the multiple images actually are sensitive to the value of $r_s$.
This can be understood as follows: the generalized NFW profile is not a power law model, in the sense that the
3D density $\rho(r)$ is changing at each radius $r$. This mass profile will be close to isothermal for
$r\sim r_s$, and the radius at which the profile becomes isothermal could be felt by observational
constraints located at $r<r_s$.
To check this scenario, we compute the 2D aperture masses inferred from each model and compare them.
If there is a significant mass difference between each model at the radius where observational constraints 
are present, then
we could understand why the observational constraints are able to discriminate between each model.
Comparison between masses, expressed as a percentage (calculated as (model(0)-model(1))/model(0)), 
is shown on Fig.~\ref{diffmass}.
We see that mass differences can reach up to 3\% at the radius where the outermost observational constraint is found
($\sim 210$\,kpc).
The question is to know whether we are sensitive to such a mass difference. In other words: is the accuracy 
on our mass measurement below this mass differences ?
From the \textsc{mcmc} realizations, we estimate the accuracy on our mass measurement and express it as a
percentage. This accuracy depends on the distance from the cluster centre, and is plotted on Fig.~\ref{diffmass}.
If mass differences are below the accuracy on our mass measurement up to $\sim$\,125\,kpc, we see that
for R$>$150\,kpc, mass differences between model (0) and models (1) and (2) are above the accuracy on
the mass measurement, which means that
we are able to discriminate between these different mass models.
This shows that the multiple images located further away than 150\,kpc from the centre are sensitive to
the value of $r_s$.

Ultimately, the scale radius of Abell~1703 should be constrained by a carefull strong and weak lensing and/or
X-ray analysis.

\begin{table*}
\begin{center}
\begin{tabular}{ccccccccc}
\hline
Clump & $\delta(x)$ & $\delta(y)$ & $e$ & $\theta$ & $r$ (\footnotesize{kpc}) & $\alpha$ & $c_{200}$ & $\sigma_0$ (\footnotesize{km\,s$^{-1}$})\\
\hline
\smallskip
\smallskip
NFW & -1.0$\pm$0.1 & 0.5$\pm$0.1 & 0.12$\pm$0.004 & 64.5$\pm$0.3& 727.0$^{+13.84}_{-76.77}$ & 1.09$^{+0.015}_{-0.036}$ &  3.1$^{+0.34}_{-0.07}$  & -- \\
\smallskip
\smallskip
\smallskip
cD  & [0.0] & [0.0] & [0.22] & [52] & [30] & -- & -- & 299.4$^{+0.35}_{-9.9}$ \\
\smallskip
\smallskip
\smallskip
Galax 852 & [19.0] & [54.0] & [0.11] & [65.5] & 97.1$\pm$1.2  & -- & --  & 319.5$\pm$4.1 \\
\smallskip
\smallskip
\smallskip
L$^*$ elliptical galaxy & -- & --& --&-- &65.9$^{+0.16}_{-4.6}$ &-- & --  & 207.8$^{+18.0}_{-15.0}$ \\
\hline
\smallskip
\end{tabular}
\end{center}
\caption{Mass model parameters.
Coordinates are given in arcseconds with respect to the cD Galaxy.
The ellipticity $e$ is the one of the mass distribution, expressed as $a^2-b^2/a^2+b^2$. Error bars correspond
to $1\sigma$ confidence level as inferred from the \textsc{mcmc} optimization. When the posterior probability
distribution is not Gaussian, we report the mode and asymmetric error bars.
Values into brackets are not optimized.
Note that the meaning of the scale radius reported here is different for the generalized NFW clump and the
other mass clumps, described using a dPIE profile with no core.
}
\end{table*}

\begin{table}
\begin{center}
\begin{tabular}{cccccc}
\hline
Limits (kpc) & $r_s$ (\footnotesize{kpc}) & $c_{200}$ & $\alpha$ & $\Delta$(log(Ev)) & $\Delta(\chi^2)$\\
\hline
\smallskip
\smallskip
(0) 150-750 &  727.0$^{+13.84}_{-76.77}$ & 3.1$^{+0.34}_{-0.07}$ & 1.09$^{+0.015}_{-0.036}$ & -  & - \\
\smallskip
\smallskip
(1) 200-300 & 299.5$^{+0.09}_{-3.28}$ & 6.9$^{+0.16}_{-0.12}$ & 0.77$^{+0.032}_{-0.021}$ & 19.6 & -41 \\ 
\smallskip
\smallskip
(2) 300-400 & 398.8$^{+0.09}_{-3.28}$ & 5.4$^{+0.20}_{-0.12}$ & 0.91$^{+0.020}_{-0.032}$ & 8.9 & -19 \\
\smallskip
\smallskip
(3) 400-600 & 598.8$^{+1.27}_{-41.18}$ & 3.7$^{+0.24}_{-0.08}$ & 1.03$^{+0.024}_{-0.021}$ & 1.9 &  -4\\
\smallskip
\smallskip
(4) 600-800 & 719.3$^{+22.94}_{-61.96}$ & 3.0$^{+0.32}_{-0.01}$ & 1.09$^{+0.019}_{-0.026}$ & -3.9 & -1 \\
\hline
\smallskip
\end{tabular}
\end{center}
\caption{Results of the analysis when using different limits for the scale radius $r_s$.
Error bars correspond to $1\sigma$ confidence level as inferred from the \textsc{mcmc} optimization.
When the posterior probability
distribution is not Gaussian, we report the mode and asymmetric error bars.
(0) corresponds to the results of the mass model presented in Table~4.
The following lines corresponds to models obtained when assuming a different prior for the scale radius.
For each model ($i$), we report the results on the parameters of the dark matter clump.
To compare the different lines, we report $\Delta$(log(Ev)) and $\Delta(\chi^2)$.
}
\end{table}

\section{Discussion}

The main result of the presented work is to measure a central slope equal to $\sim$ -1.1.
This is close to the predictions from \citet{nfw}.
However, we want to stress again that comparing these values is not relevant since dark matter only
simulations, by definition, do not take into account the likely influence from the baryonic component
on the shape of the underlying dark matter, thus
their predictions cannot be reliably compared to what is inferred observationally.

Similar analyses (i.e. lensing analyses aiming to probe the central mass density
distribution in galaxy clusters) have been carried out by \citet{sand02,sand04,sand07}
\citep[see also][on MS\,2137-23]{gavazzi03,gavazzi05}.
Studies by Sand et~al. used the measured velocity dispersion profile of the cD galaxy as an extra constrain.
The first work by \citet{sand04} assumed a circular cluster.
As shown by \citet{massimo07}, this assumption was likely to bias the results towards shallower
values of the central mass density slope.
Then \citet{sand07} redid their analysis using a full 2D lensing analysis, taking into account the
presence of substructures and allowing the clusters for non circularity.
They confirmed their earlier claims, in particular, they do find evidence for a central mass density slope to 
be smaller than 1.
The results presented in this work points out towards higher values for the slope of the
central mass distribution compared to the studies by Sand et~al.
It is worth mentioning that our analysis is close to the ones by Sand et~al., since we have both
used strong lensing techniques, and moreover used almost the same \textsc{lenstool} software 
(by the time of the studies by Sand et~al., the Bayesian Monte Carlo Markov Chain sampler as described
by \citet{jullo07} was not available, and they used a parabolic $\chi^2$ optimization instead, but
this cannot explain the differences).
This wide range of slopes found from one cluster to another may point out to an intrinsic large scatter
on this parameter, which may depend on the merger history of the cluster. This possible scatter should be probed numerically.

\paragraph{Prospects on Abell~1703:}
We can summarize some possible extensions of the presented work:
\begin{enumerate}
\item We have left some blue features that are likely to be lensed. Therefore, the multiply imaged systems 
identification is still to be improved. 
In particular, it would be useful to understand the radial feature that is coming out from the 
cD galaxy.
\item Parametric strong lensing is very sensitive to misidentifications, thus we are pretty confident of our
identifications. However, a measure of their redshifts would be very valuable.
\item Probing the cluster potential on larger scales: our analysis allowed the scale radius to 
vary within a large prior suggested by numerical simulations. We have investigated how the choice
of this prior can influence the results on $\alpha$, finding substantial differences
between different priors. Measuring reliably the scale radius is thus of first importance and should be
done combining strong and weak lensing and/or X-ray measurements.
\item Measuring the velocity dispersion of the stars of the cD galaxy in order to add dynamical constraints
as in the studies by Sand et~al.
\item The more relaxed Abell~1703, the stronger the conclusions of the work presented here, and also
the easier the comparison with N-body simulations.
We have argued above that we find evidence for a relaxed cluster.
It would be important to check further the dynamical state of Abell~1703.
In particular, high resolution X-ray observations of Abell~1703 could help to answer this question.
Moreover, X-ray observations would also make possible to subtract the gas contribution from the mass budget, following
\citet{bradac}.
\item Moreover, the spectroscopic study of the motions of cluster galaxies provides clues to the dynamical state
of the cluster. Such data give an insight into the velocity dimension, and can reveal if the cluster
is undergoing a merger along the line of sight or if it is already well relaxed.
Of particular interest will be to probe the velocity difference between the cD galaxy and the brightest cluster
galaxies defining this filamentary structure, in order to investigate if we are observing an infalling
galaxy group.
\item Related to this group issue is the study of Galaxy 852 based on the detection of a nearby arc 
and the surrounding multiply imaged systems.
\end{enumerate}

\paragraph{Prospects on the Central Mass Distribution of Galaxy Clusters:}
Obviously, we suffer from small number statistics and we need to extend this kind of analyses to other 
clusters.
In trying to gather a sample of clusters for which the presented analysis could be applied, one should
focuss on relaxed unimodal cD dominated clusters which present observational constraints as close as possible
to the centre of the cluster. In that respect, one should look for radial arcs.
Note, however, that radial arcs might preferentially form in clusters with shallow
profile. Ideally, selection should be independent of arc appearances, but the clusters with
radial arcs will indeed give good constraints.

On the numerical side, we need to study a sample of many galaxy clusters containing baryons, and 
testing different prescriptions for the baryonic implementation.
We have initiated such a numerical study on two galaxy clusters, and results will be presented in a
forthcoming publication.

Conducting in parallel an observational and a numerical program is a worthy goal:
it is interesting by itself to study what is going on in the central part of the most massive virialized
structures since it can provide insights on the interactions between baryons and dark matter particles;
moreover, it can potentially provide an interesting probe of cosmological models.

\section*{Acknowledgement}
We thank many people for constructive comments and discussion related to this topic, in particular:
Bernard Fort, Dave Sand, Hans B{\"o}hringer, Jens Hjorth, Kristian Pedersen, Steen Hansen. 
We thank John Stott for creating the colour image from which Fig.~1 has been made, and for allowing
us to use it.
The referee is acknowledged for a careful reading and a constructive report.
ML aknowledges the Agence Nationale de la Recherche for its support.
The Dark Cosmology Center is funded by the Danish National Research Foundation.
JR is grateful to Caltech for its support.
JPK aknowledges the Centre National de la Recherche Scientifique for its support, Project number BLAN06-3-135448.
We thank the Danish Centre for Scientific Computing at the University of Copenhagen for
providing us generous amount of time on its supercomputing facility.
Based on data collected at Subaru Telescope, which is operated by the National Astronomical Observatory of Japan.
We are thankful to Ichi Tanaka for his support in the reduction of MOIRCS imaging data.
ML aknowledge the lensing group at Shangai  Normal University for their kind invitation and
hospitality, during which this work has been initiated.
The authors recognize and acknowledge the very significant cultural
role and reverence that the summit of Mauna Kea has always had
within the indigenous Hawaiian community.  We are most fortunate
to have the opportunity to conduct observations from this mountain.

\bibliographystyle{aa} 
\bibliography{draft}

\include{appendix}

\end{document}

%% file: table.tex
\begin{table*}
\begin{center}
\begin{tabular}{ccccccccc}
\smallskip
Id & R.A. & Decl. & $z_{\rm phot}$ &  $\Delta(z_{\rm phot})$ & $z_{\rm mod}$ & RMS (s) & RMS (i) & $z_{\rm free}$\\
\hline
\hline
\smallskip
1.1 & 198.77725 & 51.81934  & 0.965$_{-0.240}^{+0.075}$   & - 0.8885 - & - & 0.27 &1.95	& \\
\smallskip
1.2 & 198.77482 & 51.81978  & 0.940$_{-0.096}^{+0.102}$   & \ldots  &	& & &	\\
\smallskip
1.3 & 198.77414 & 51.81819  & 0.995$_{-0.087}^{+0.051}$   & \ldots 	& & &	&	\\
\smallskip
1.4 & 198.77671 & 51.81767  & 0.740$_{-0.078}^{+0.129}$   & \ldots 	& & &	&	\\
\smallskip
1.5 & 198.76206 & 51.81331  & 0.915$_{-0.090}^{+0.108}$   & \ldots &	& & &	\\
\smallskip
2.2 & 198.77156 & 51.81174  & 2.310$_{-0.381}^{+0.456}$  &  [1.9 - 2.8] & 2.24$_{-0.14}^{+0.25}$ & 0.09 &0.39 &  3.98$_{-0.52}^{+1.30}$ \\
\smallskip
2.3 & 198.76970 & 51.81186  & 2.225$_{-0.657}^{+0.195}$  & \ldots 	&	& & &	\\
\smallskip
3.1 & 198.76696 & 51.83205  &    ----                  & [3.31-3.37] &	3.32$_{-0.01}^{+0.03}$ & 0.15	&1.22 &	4.85$_{-0.71}^{+0.65}$ \\
\smallskip
3.2 & 198.76634 & 51.83190  & 3.350$_{-0.036}^{+0.024}$  & \ldots &	& &  &		\\
\smallskip
3.3 & 198.75824 & 51.82982  & 3.350$_{-0.036}^{+0.024}$  & \ldots &	& &  &		\\
\smallskip
4.1 & 198.76564 & 51.82653  &           ----           & [2.03-2.48] &  2.03$_{-0.00}^{+0.03}$ &0.17&0.51&2.46$_{-0.11}^{+0.12}$\\
\smallskip
4.2 & 198.76075 & 51.82487  & 2.255$_{-0.222}^{+0.216}$  &\ldots  &	& & &	\\
\smallskip
4.3 & 198.77666 & 51.82797  &         ----             &\ldots  &	& & &	\\
\smallskip
5.1 & 198.76602 & 51.82677  &       ----                & [2.03-2.48] & 2.03$_{-0.00}^{+0.03}$& 0.21	&0.64&	2.36$_{-0.08}^{+0.14}$\\
\smallskip
5.2 & 198.76041 & 51.82489  & 2.315$_{-0.186}^{+0.171}$   &\ldots  &	&  &  &	\\
\smallskip
5.3 & 198.77591 & 51.82807  &      ----                    & \ldots &	& &  &	\\
\smallskip
6.1 & 198.77984 & 51.82640  & 2.595$_{-0.159}^{+0.144}$   & [2.43-2.79] &  2.78$_{-0.06}^{+0.00}$& 0.30	&0.84 &	4.98$_{-0.49}^{+0.45}$\\
\smallskip
6.2 & 198.76890 & 51.82580  & 2.535$_{-0.117}^{+0.258}$   & \ldots &	& & &	\\
\smallskip
6.3 & 198.75652 & 51.81947  & 2.625$_{-0.135}^{+0.108}$   & \ldots &	& & &	\\
\smallskip
7.1 & 198.77074 & 51.83087  & 3.490$_{-0.108}^{+0.132}$   & [2.52-3.62] & 3.59$_{-0.18}^{+0.02}$& 0.28&2.59 &	5.63$_{-0.56}^{+0.02}$\\
\smallskip
7.2 & 198.76614 & 51.83010  & 2.960$_{-0.183}^{+0.354}$   & \ldots &	& & &	\\
\smallskip
7.3 & 198.75869 & 51.82814  & 3.200$_{-0.675}^{+0.210}$  &\ldots  &	& & &	\\
\smallskip
8.1 & 198.77250 & 51.83045  & 2.805$_{-0.090}^{+0.174}$  & [2.61-2.98] &  2.97$_{-0.05}^{+0.00}$ &0.30	&1.63&	 5.53$_{-0.61}^{+0.08}$\\
\smallskip
8.2 & 198.76608 & 51.82949  & 2.770$_{-0.108}^{+0.207}$  & \ldots &	& & &	\\
\smallskip
8.3 & 198.75863 & 51.82740  & 2.725$_{-0.117}^{+0.198}$  & \ldots &	& & &	\\
\smallskip
9.1 & 198.77176 & 51.83030  &  ----                      & [2.40-3.37] & 3.36$_{-0.07}^{+0.04}$ & 0.24	&1.29&5.53$_{-0.61}^{+0.08}$	\\
\smallskip
9.2 & 198.76690 & 51.82957  & 2.995$_{-0.378}^{+0.195}$  & \ldots &	& & &	\\
\smallskip
9.3 & 198.75813 & 51.82708  & 3.000$_{-0.603}^{+0.366}$  & \ldots &	& & &	\\
\smallskip
10.1 & 198.78708 & 51.81424 & 3.100$_{-0.162}^{+0.324}$  & [2.40-3.42] & 2.41$_{-0.01}^{+0.07}$ & 0.21	&1.62 &3.79$_{-0.28}^{+0.25}$	\\
\smallskip
10.2 & 198.78352 & 51.81138 & 2.595$_{-0.189}^{+0.117}$  & \ldots &	& & &	\\
\smallskip
10.3 & 198.76242 & 51.80954 & 2.705$_{-0.261}^{+0.189}$  & \ldots &	& & &	\\
\smallskip
11.1 & 198.78648 & 51.81322 & 3.045$_{-0.138}^{+0.102}$  &  [2.40-3.42] & 2.47$_{-0.04}^{+0.06}$ & 0.25	&2.13&	3.79$_{-0.27}^{+0.27}$\\
\smallskip
11.2 & 198.78564 & 51.81247 & 3.155$_{-0.099}^{+0.078}$  & \ldots &	& & &	\\
\smallskip
11.3 & 198.76242 & 51.80954  & 2.705$_{-0.261}^{+0.189}$  & \ldots &	& & &	\\
\smallskip
15.1 & 198.76284 & 51.81246  & 2.440$_{-0.231}^{+0.171}$   &  [2.21-2.67]& 2.67$_{-0.09}^{+0.00}$& 0.37	&0.93&5.58$_{-0.71}^{+0.26}$	\\
\smallskip
15.2 & 198.76704 & 51.82128  & 2.440$_{-0.153}^{+0.231}$  & \ldots  &	& & &	\\
\smallskip
15.3 & 198.78821 & 51.82176  &   ----                     & \ldots  &	& & &	\\
\smallskip
15.4 & 198.77519 & 51.81155  &   -----                  & \ldots  &	& & &	\\
\smallskip
16.1 & 198.76356 & 51.81164 & 2.710$_{-0.324}^{+0.216}$  & [2.50 - 2.92] & 2.70$_{-0.08}^{+0.10}$& 0.27	&0.73&	5.04$_{-0.52}^{+0.44}$\\
\smallskip
16.2 & 198.76774 & 51.82101 &  ----           & \ldots   &	& & &		\\
\smallskip
16.3 & 198.78838 & 51.82097 & 2.750$_{-0.165}^{+0.171}$  & \ldots   &	& & &	\\
\smallskip
16.4 & 198.77558 & 51.81128 &    ----   & \ldots   &	& & &	\\
\smallskip
\end{tabular}
\end{center}
\label{multipletable}
\caption{Multiply imaged systems considered in this work. We have found 13 distinct multiply imaged
systems.
Coordinates are given in degrees (J\,2000.0).
When the photometry is reliable in each band, we report the corresponding photometric redshift estimates, 
with error bars quoting the 3$\sigma$ confidence level.
$\Delta(z_{\rm phot})$ corresponds to the redshift range allowed by the photometric redshift estimation and 
that will be used as a prior
in the optimization (i.e. for each system, the redshift will be let free and allowed to vary between 
$\Delta(z_{\rm phot})$).
For system 1 however we fix the redshift to the measured one, 0.8885. The photometric
estimate $\Delta(z_{\rm phot})$\,=\,[0.725 - 1.04] is in agreement with the spectroscopic measurement. 
$z_{\rm mod}$ corresponds to the redshift inferred from the optimization procedure.
We report both the RMS in the source plane and the RMS in the image plane.
The mean scatters are given for the whole system, not for each individual image composing a system.
The total RMS is equal to 0.26$\arcsec$ (source plane) and 1.45$\arcsec$ (image plane).
$z_{\rm free}$ corresponds to the redshift inferred from the optimization when all redshift but system 1 are assigned a
flat prior between 0.28 and 6 (see Section 5.3).
}
\end{table*}

%% file: appendix.tex
\appendix
\section{Multiply Imaged systems}
In this Appendix we show colour pictures of the multiply imaged systems used in this work.
The size of each panel is indicated in the caption.
All images are aligned with the WCS coordinates, i.e. north is Up, east is Left.

\begin{figure}
\centering
\includegraphics[height=8.5cm,width=8.5cm]{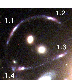}
\includegraphics[height=8.5cm,width=8.5cm]{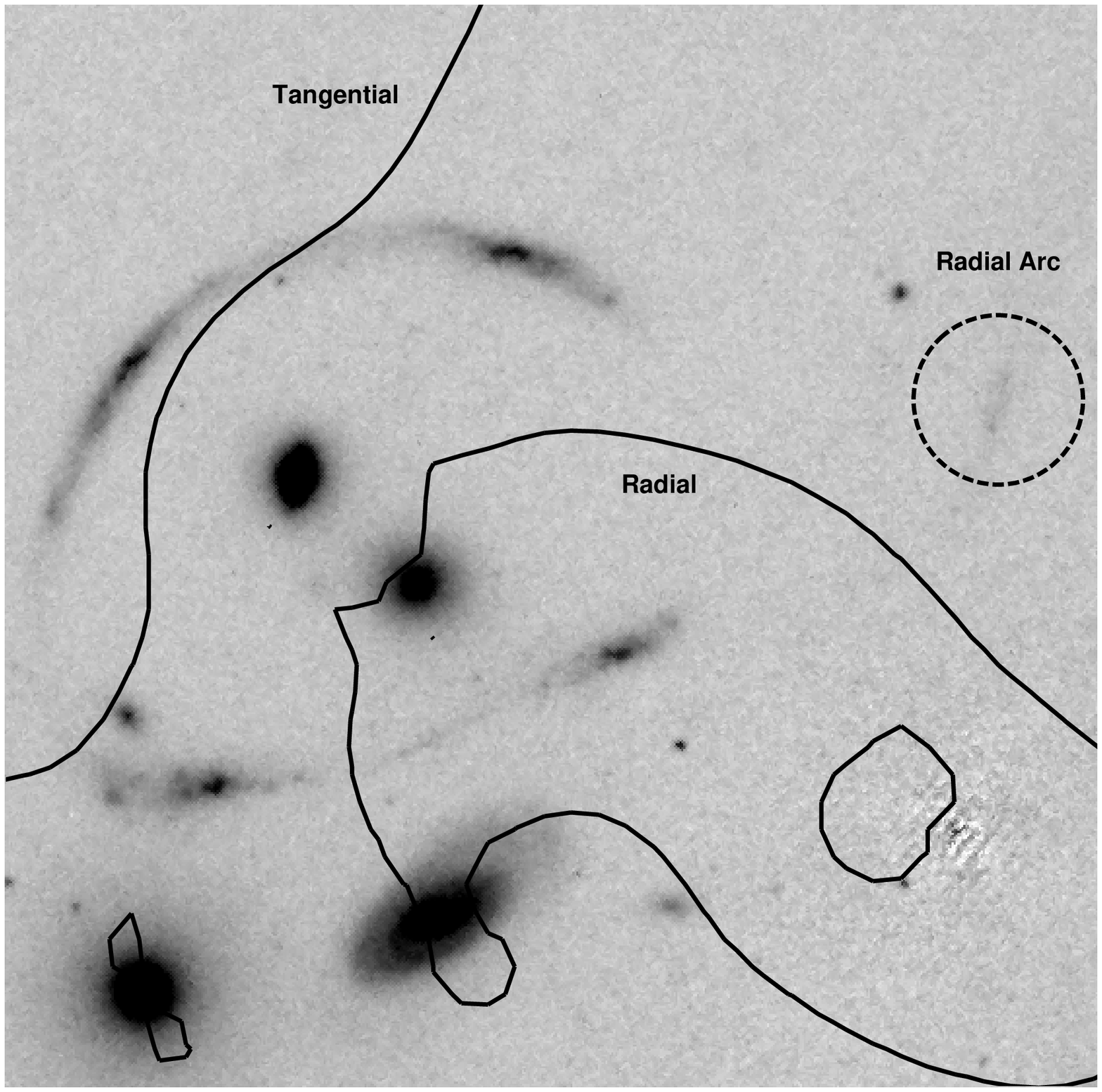}
\caption{
System 1, the \emph{central ring} composed by 4 bright images.
For its counter image, see Fig.~\ref{sys15-16} below.
\emph{Top panel:} colour image from F850W, F625W and F465W observations.
Size of panel is $10\arcsec\,\times\,10\arcsec$.
\emph{Botom panel:} F775W image where the light from the cD galaxy has been subtracted.
We plot the tangential and critical
	lines at $z=0.88$ (redshift of system 1).
	We also report a radial feature that is coming out from the cD galaxy (see Fig.~1), that is
	not used in the analysis since we have not been able to detect the counter images.
	Size of panel is $16\arcsec\,\times\,16\arcsec$.
	}
\label{sys1}
\end{figure}

\begin{figure}
\centering
\includegraphics[height=8.5cm,width=8.5cm]{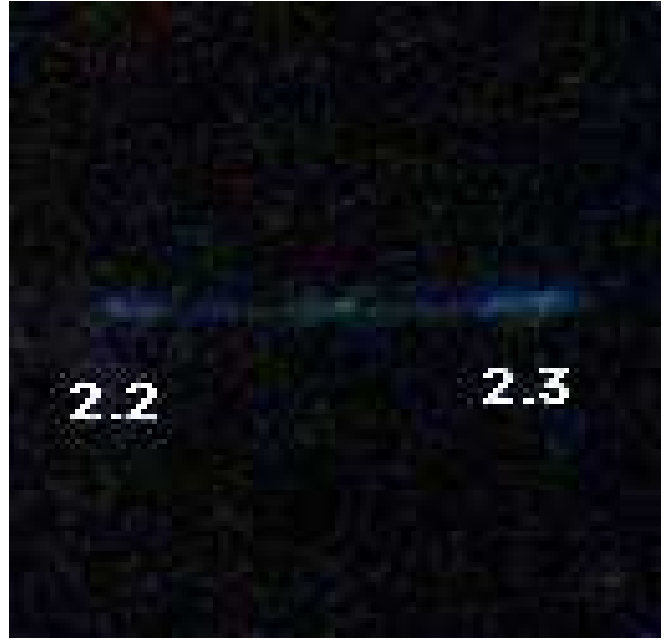}
\caption{
System 2, constituted by two magnified merging images. Two counter images are predicted to be more than
two magnitudes fainter, we have not been able to detect any.
Size of panel is $6.3\arcsec\,\times\,6.3\arcsec$.
}
\label{sys2}
\end{figure}

\begin{figure}
\centering
\includegraphics[height=8.5cm,width=8.5cm]{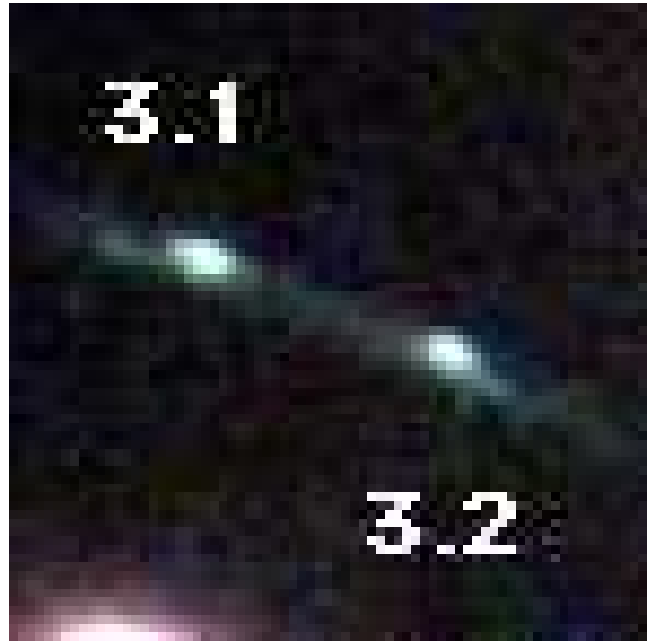}
\caption{
System 3, constituted by two magnified merging images and a demagnified one a bit further west.
Size of panel is $4\arcsec\,\times\,4\arcsec$.
}
\label{sys3}
\end{figure}
\begin{figure}
\centering
\includegraphics[height=8.5cm,width=8.5cm]{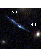}
\includegraphics[height=8.5cm,width=8.5cm]{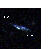}
\includegraphics[height=8.5cm,width=8.5cm]{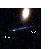}
\caption{
System 4-5.
These two systems are assumed to belong to the \emph{same} background source galaxy.
System 4 is systematically brighter than system 5.
{\emph{From top to bottom:}} images 4.1 and 5.1; 4.2 and 5.2;  4.3 and 5.3.
Size of each panel is $6.5\arcsec\,\times\,6.5\arcsec$.
}
\label{sys4-5}
\end{figure}

\begin{figure}
\centering
\includegraphics[height=8.5cm,width=8.5cm]{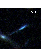}
\includegraphics[height=8.5cm,width=8.5cm]{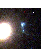}
\includegraphics[height=8.5cm,width=8.5cm]{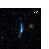}
\caption{
System 6. {\emph{From top to bottom:}} image 6.1, 6.2 and 6.3.
Size of each panel is $6.5\arcsec\,\times\,6.5\arcsec$.
}
\label{sys6}
\end{figure}

\begin{figure}
\centering
\includegraphics[height=8.5cm,width=8.5cm]{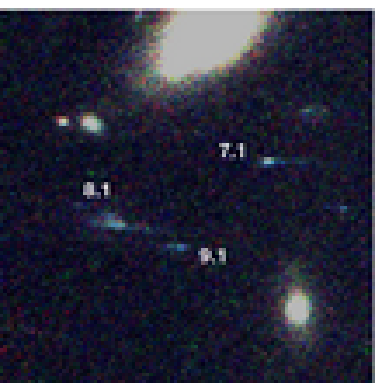}
\includegraphics[height=8.5cm,width=8.5cm]{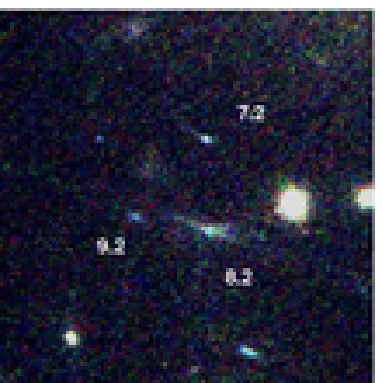}
\includegraphics[height=8.5cm,width=8.5cm]{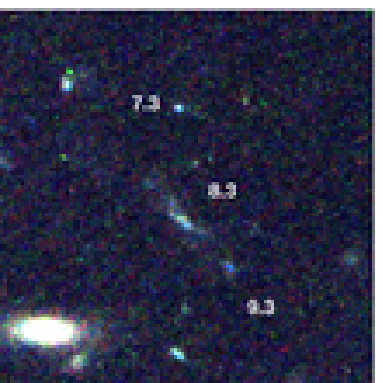}
\caption{
Systems 7, 8 and 9. Size of each panel is $8.5\arcsec\,\times\,8.5\arcsec$.}
\label{sys7-8-9}
\end{figure}

\begin{figure}
\centering
\includegraphics[height=11.1cm,width=8.5cm]{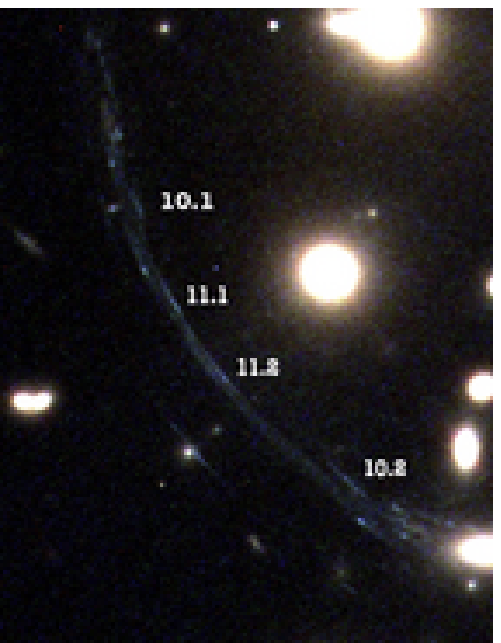}
\includegraphics[height=8.5cm,width=8.5cm]{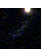}
\caption{
 Systems 10 and 11 corresponds to two substructures identified on the giant arc.
 Note than more can be defined along this giant arc.
 \emph{Top panel:} images 10.1, 11.1, 11.2 and 10.2. Size of panel is $18\arcsec\,\times\,12\arcsec$.
 \emph{Bottom panel:}
 counter image of both systems, also presenting substructures. Size of panel is $6.6\arcsec\,\times\,6.6\arcsec$.
    }
\label{sys10-11}
\end{figure}
\clearpage

\begin{figure*}
\centering
\includegraphics[height=8.5cm,width=8.5cm]{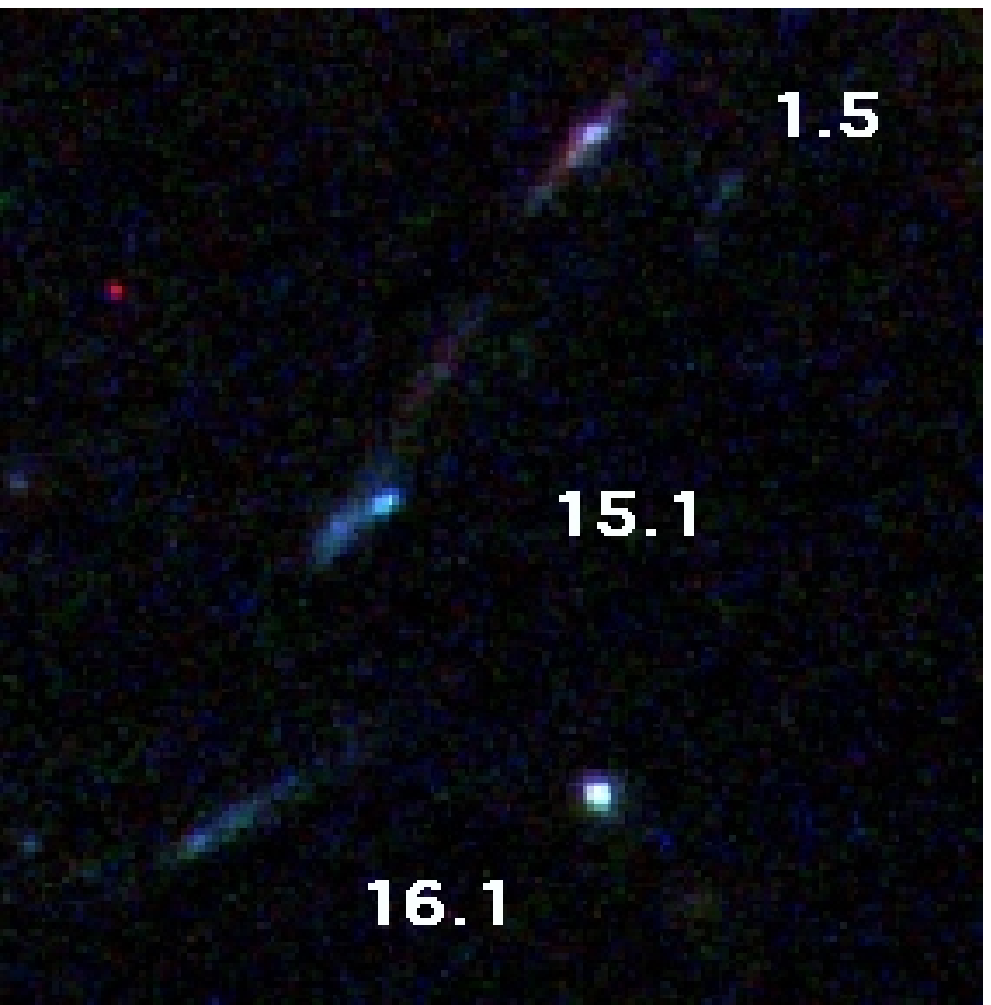}
\includegraphics[height=8.5cm,width=8.5cm]{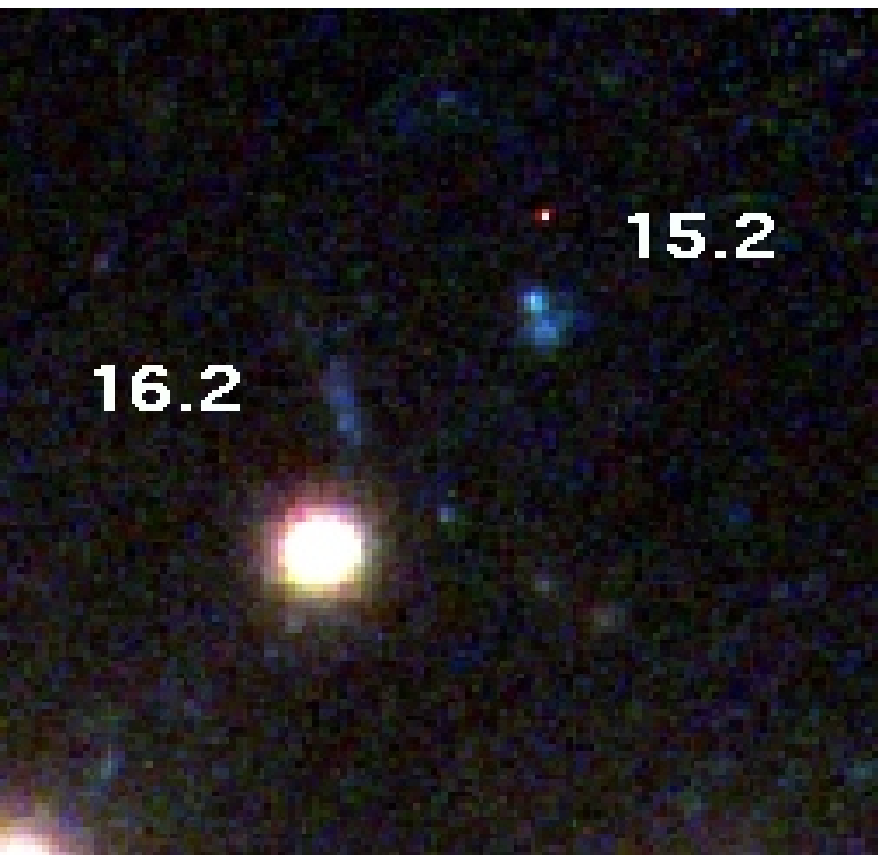}\\
\includegraphics[height=8.5cm,width=8.5cm]{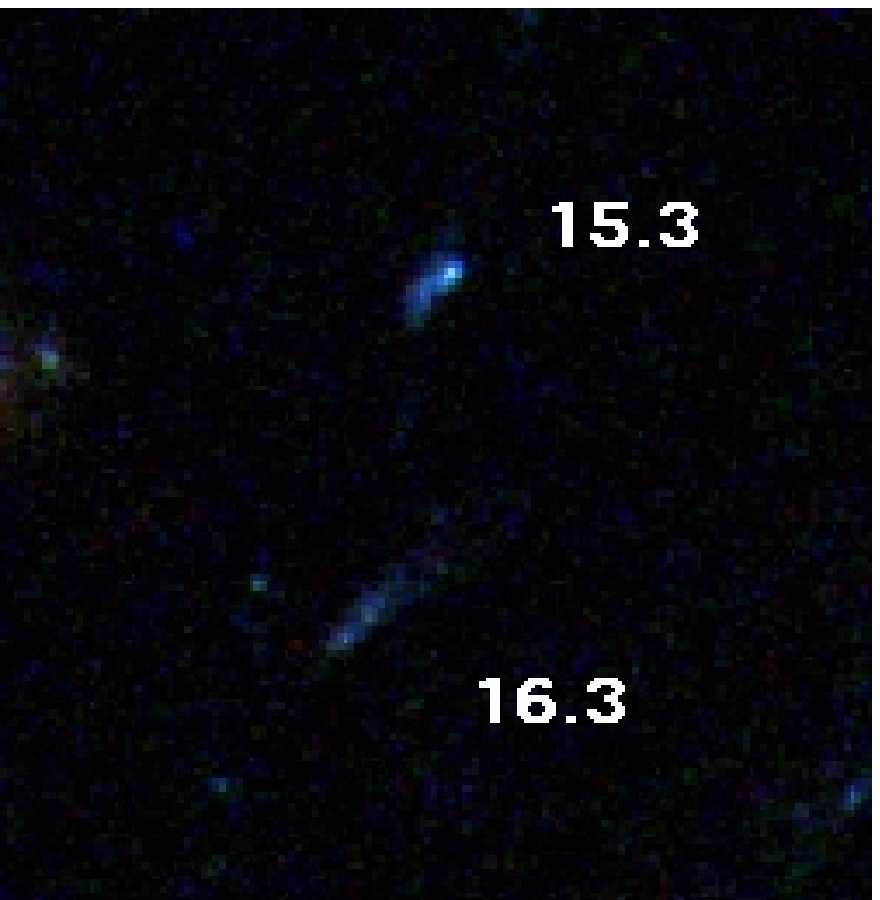}
\includegraphics[height=8.5cm,width=8.5cm]{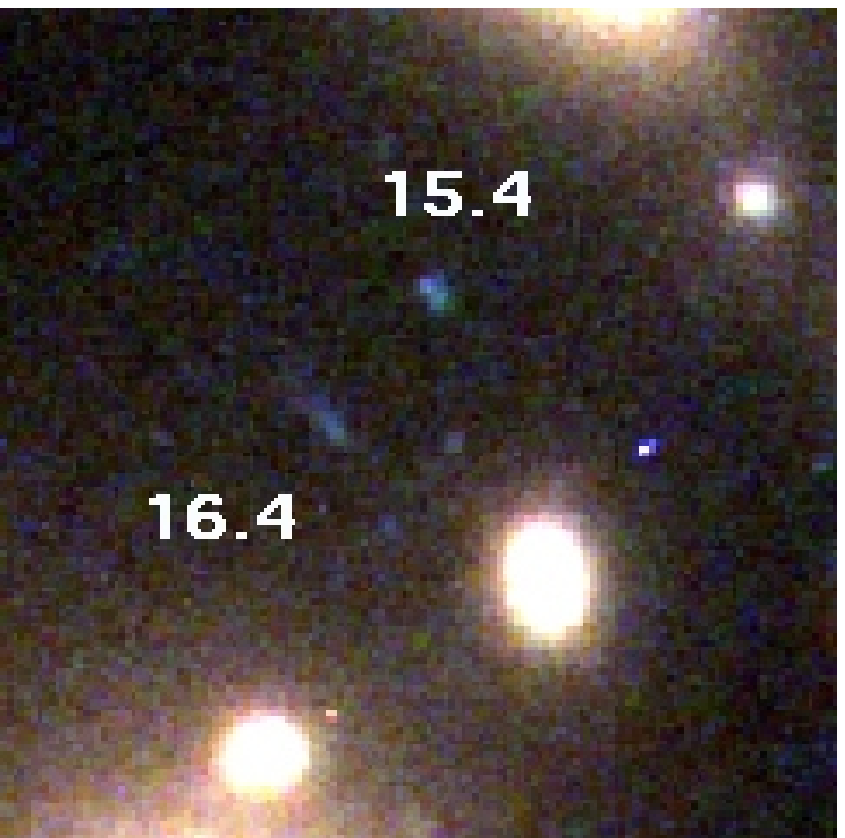}
\caption{
Systems 15 and 16.
Images belonging to system 15 are systematically brighter than images belonging to system 16.
\emph{First panel}, north to south: image 1.5, counter image of the \emph{central ring} (Fig.~\ref{sys1}); 
image 15.1 and image 16.1.
\emph{Second panel:} images 15.2 and 16.2. 
\emph{Third panel:} images 15.3 and 16.3.
\emph{Last panel:} images 15.4 and 16.4. Size of each panel is $8\arcsec\,\times\,8\arcsec$.
    }
\label{sys15-16}
\end{figure*}
\clearpage
\begin{figure}
\centering
\includegraphics[height=8.5cm,width=8.5cm]{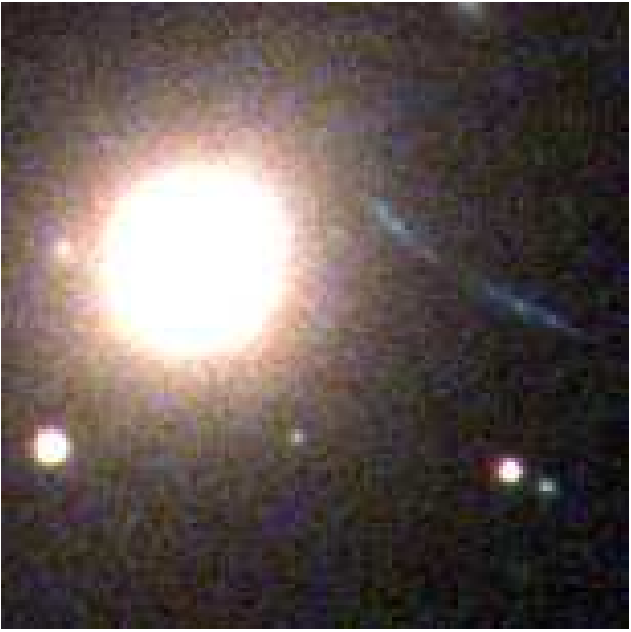}
\caption{
Galaxy 852, located in the northern part of the ACS field, at $\alpha=198.76342; \delta=51.832523$.
Size of panel is $8.2\arcsec\,\times\,8.2\arcsec$.
The blue radial feature is easy to detect on this picture.
}
\label{G852}
\end{figure}

\clearpage
\newpage
\begin{figure*}
\centering
\includegraphics[height=25cm,width=15cm]{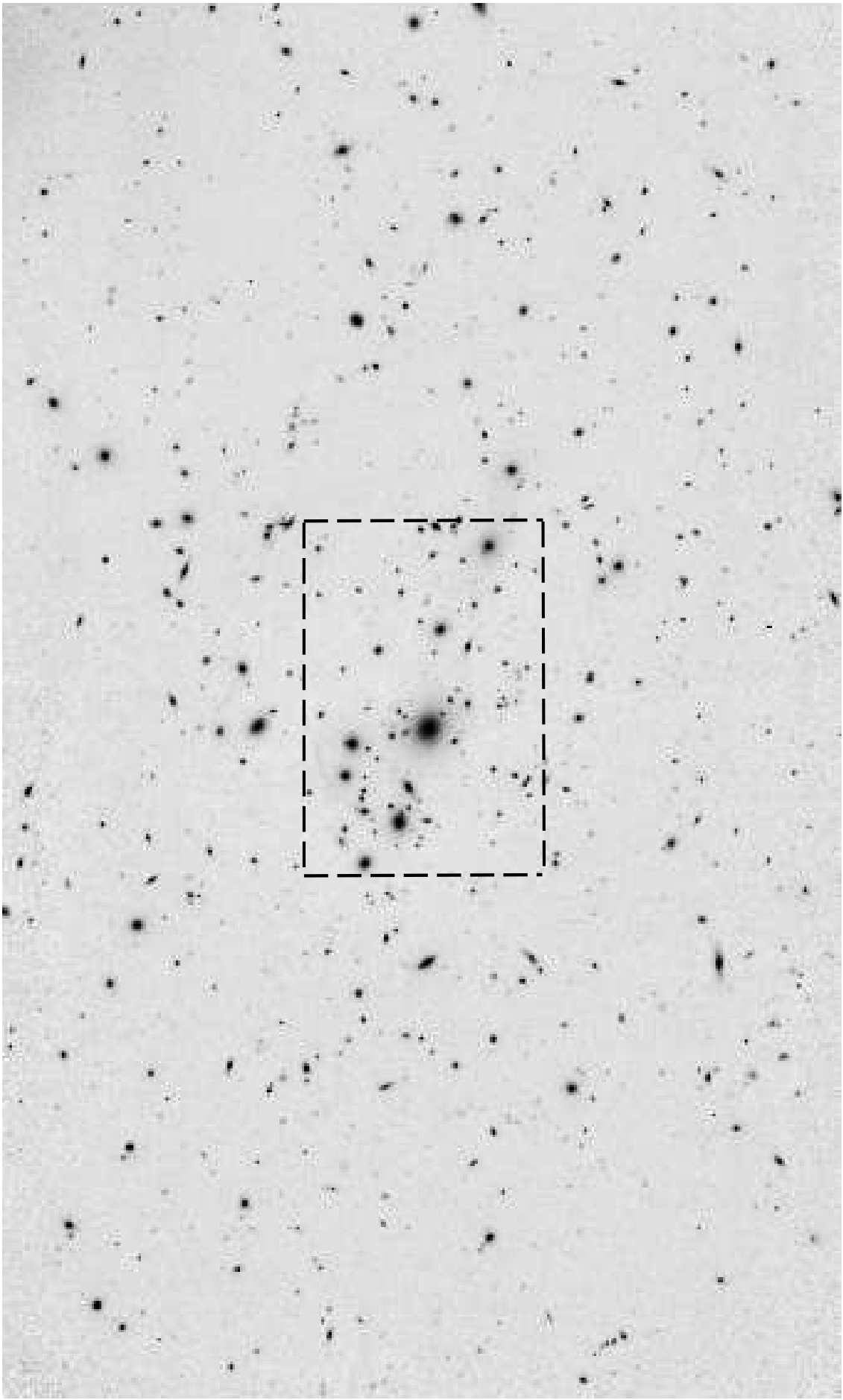}
\caption{
Subaru image of Abell~1703 (H band). Size of panel is $275\arcsec\,\times\,452\arcsec$.
The filamentary structure identified on Fig.~1 can be appreciated further on this image.
The frame corresponds to the size of Fig.~1.
}
\label{subaru}
\end{figure*}